\documentclass[useAMS]{mn2e}

\usepackage{rotating}
\usepackage{lscape}
\usepackage{subfigure}
\usepackage{graphicx}
\usepackage{longtable}
\usepackage{float}
\usepackage{captcont}
\usepackage{endfloat}
\usepackage{morefloats}

\title[Galaxies with Background QSOs, I]{Galaxies with Background QSOs, I: A Search for Strong Galactic H-alpha Lines}

\author[Donald G. York et al.]
{Donald G. York,$^1$\thanks{Also, the Enrico Fermi Institute.} Lorrie A. Straka,$^2$ Michael Bishof,$^1$ Seth Kuttruff,$^3$ 
\newauthor
David Bowen,$^4$ Varsha P. Kulkarni,$^2$ Mark Subbarao,$^5$ Gordon Richards,$^6$  
\newauthor
Daniel Vanden Berk,$^7$ Patrick B. Hall,$^8$ Timothy Heckman,$^9$ Pushpa Khare,$^{10}$  
\newauthor
Jean Quashnock,$^{3,1}$ Lara Ghering,$^1$ Sean Johnson$^1$\\
$^1$Department of Astronomy and Astrophysics, University of Chicago, Chicago, IL 60637\\
$^2$Department of Physics and Astronomy, University of South Carolina, Columbia, SC 29208\\
$^3$Department of Physics and Astronomy, Carthage College, Kenosha, WI 53140\\
$^4$Department of Physics and Astronomy, Princeton University, Princeton, NJ, 08544\\
$^5$Adler Planetarium, Chicago, IL 60605\\
$^6$Department of Physics, Drexel University, Philadelphia, PA 19104\\
$^7$Department of Physics, St. Vincent College, Latrobe, PA 15650\\
$^8$Department of Physics and Astronomy, York University, Toronto, ON, Canada\\
$^9$Department of Physics and Astronomy, Johns Hopkins University, Baltimore, MD 21218\\
$^{10}$CSIR Emeritus Scientist, IUCAA, Ganeshkhind, Pune, India 411007\\
}

\date{}

\pagerange{\pageref{firstpage}--\pageref{lastpage}} \pubyear{}

\def\LaTeX{L\kern-.36em\raise.3ex\hbox{a}\kern-.15em
    T\kern-.1667em\lower.7ex\hbox{E}\kern-.125emX}

\begin{document}

\maketitle

\label{firstpage}

\begin{abstract}

A search for emission lines in foreground galaxies in quasar spectra ($z_{gal} < z_{QSO}$) of the Sloan Digital Sky Survey (SDSS) data release 5 (DR5) reveals 23 examples of quasars shining through low redshift, foreground galaxies at small impact parameters ($<10$ kpc).  About 74,000 quasar spectra were examined by searching for narrow H$\alpha$ emission lines at $z < 0.38$, at a flux level greater than $5 \times 10^{-17}$ ergs cm$^{-2}$ s$^{-1}$, then confirming that other expected emission lines of the H II regions in the galaxy are detected.  The galaxies were deblended from the quasar images to get colors and morphologies. For cases that allow the galaxy and the quasar to be deblended, the galaxies are blue ($0.95<$(u-r)$<1.95)$. Extinction and reddening through the galaxies is determined from the (g-i) color excesses of the quasars. These reddening values are compared with the flux ratio of H$\alpha$ to H$\beta$, which reflect the extinction for an undetermined fraction of the sightline through each galaxy. No trends were found relating E(B-V)$_{(g-i)}$, impact parameter (b), and (u-r) for the galaxies or between E(B-V) derived from (g-i) and that derived from H$\alpha$/H$\beta$. Comparison with previous studies of quasar absorption systems indicate our sample is more reddened, suggesting disk-dominated absorber galaxies. Measurement or limits on galactic, interstellar Ca II and Na I absorption lines are given from the quasar spectrum. No trends were found relating Ca II equivalent width (W (Ca II)) or Na I equivalent width (W (Na I)) to b, but a correlation of r$_{s}=-0.77$ ($\alpha=0.05$) was found relating W (Ca II) and E(B-V)$_{(g-i)}$.

\end{abstract}


\section{Introduction}

Since the earliest days of obtaining quasar spectra, it has been known that systems of intervening absorption lines appear in the spectra of quasars ($z_{abs} < z_{em}$) (Sandage 1965, Burbidge et al. 1966, Schmidt 1966, Stockton \& Lynds 1966, Arp et al. 1967). Absorption line systems seen in quasar spectra (QSOALS) with apparent velocities with respect to the quasar of $\beta > 0.02$ (where $\beta=$ v/c) are most often interpreted as arising in intervening galaxies (Bahcall et al. 1966, Bahcall 1968, Bahcall \& Spitzer 1969, Sargent et al. 1989, Steidel et al. 1995). In some cases, these absorbers may arise in ejected material associated with the background quasar, but appearing at velocities up to 30,000 km s$^{-1}$ (Hamman et al. 1997, Richards et al. 1999, Misawa et al. 2007).  However, we are not presently concerned with systems associated with the quasar. 

Of chief concern here are those systems with $\beta > 0.02$. The issue is important for several reasons: 1) Absorption line systems arising in quasar spectra provide a straightforward and precise method of probing abundances in galaxies in the range $0<z<6$, and hence the general evolution of the elements (Pettini et al. 1997, Kulkarni et al. 2007, Prochaska et al. 2006), 2) The distribution and history of gas in galaxies, which may extend to distances up to $0.5$ Mpc (Steidel et al. 1995), can be probed (Heckman et al. 2001, Adelberger et al. 2003, Adelberger et al. 2005, Tumlinson \& Fang 2005, Kenney et al. 2006, Pontzen et al. 2008, Tumlinson et al. 2011).  3) The gas distribution in galaxies may be related to the origin of analogous 21cm, high velocity cloud systems of the Milky Way (Wakker et al. 2007, Wakker \& van Woerden 1997) which systems presumably exist in other galaxies (Appleton et al. 1981, York 1982, Pisano et al. 2007, Osterloo et al. 2007). 4) Studies of a large sample of low- and high-z absorbers associated with galaxies may provide information on the amount of gas and dust in  galaxies of different ages and morphologies. 5) Comparison of the properties of those absorbers with known galaxies to absorbers for which no galaxies are seen may reveal a way to separate high velocity ejecta originating in the background quasar from absorption in foreground, but invisible, galaxies.

Until recently, the galaxies associated with high $\beta$ systems have been hard to detect via imaging, and the origin of many absorbers could not be confirmed (e.g. Cohen et al. 1987, Boisse \& Bergeron 1987, Yanny \& York 1992, Bechtold \& Ellingson 1992, Le Brun et al. 1997, Bouche et al. 2001, Chen et al. 2005, Kulkarni et al. 2005, Kulkarni et al. 2006, Gharanfoli et al. 2007).  Fewer than 60 such galaxies are confirmed (out to $\sim50$ kpc; Chen et al. 2010, P\'eroux et al. 2011a,b,  P\'eroux et al. 2012), with a further $\sim 100$ unconfirmed but still possible candidates (Straka et al. 2010, 2011;  Rao et al. 2011; Meiring et al. 2011; P\'eroux et al. 2011a,b).  Recent progress at higher redshifts (z$>$2) reveals several confirmed detections (Moller et al. 2002, 2004; Fynbo et al. 2010, 2011; Bouche et al. 2012; Noterdaeme et al. 2012; Kulkarni et al. 2012). This is a sample too small to establish a complete, unique set of properties of absorption systems from galaxies and their halos (Womble et al. 1993).  We have here begun to expand the sample with a preliminary pool of 23 fields, 10 of which are unique. In our subsequent papers, we will be adding a further $>50$ unique fields to the overall sample, allowing us to make a more complete analysis.

In order to recognize foreground galaxies associated with QSOALS, we must first establish their properties. One way to establish these properties is to carry out absorption studies in a very large number of low-z galaxies ($z<0.5$), for which certain galaxy properties are easy to determine (mass, stellar population age, and morphology, for example). Given that we have an excellent statistical characterization of QSOALS at high-z (Sargent et al. 1989, Steidel \& Hamilton 1992, LeBrun et al. 1997, Churchill et al. 2003, York et al. 2006, Khare et al. 2007, Vanden Berk et al. 2008), we need a significant data set of absorption line systems in known low-z galaxies.  The addition of a low-z sample will form a more complete evolutionary picture over a range of redshifts when combined with high-z samples. Though the sample of low-z absorbers will in all likelihood not translate directly to the high-z sample due to this evolution of properties, we may be able to discern, using analogous properties, those QSOALS at high-z that arise in foreground galaxies. We can extend the argument in order to understand in which subparts of galaxies those absorption features arise (disks, systems of 21cm high velocity clouds, tidal streams, etc.), and also to understand key physical properties such as the sources of ionization (C IV, O VI) within galaxies (Bowen et al. 2008). It would likewise be important to learn to distinguish ejected systems, spectroscopically,  from intervening systems  (Hamman et al. 1997, Hall et al. 2002, Reichart et al. 2004, Trump et al. 2006) for further study of those entities, but that is beyond the scope of this paper.

One way to work around the fact that many searches for absorbers have found no counterpart galaxy is to observe visible or spectroscopically detected galaxies on top of QSO images. A number of ways exist to find quasar-galaxy pairs (QGP). Pairs of quasars and galaxies have been most often found by matching catalogs of quasars and galaxies (Hewitt \& Burbidge 1989, Womble et al. 1993,  Bowen et al. 2006) or by first finding absorption lines in quasar spectra and following up with deep imaging to find nearby galaxies with matching redshifts (Bergeron \& Boisse 1991, Yanny \& York 1992, Le Brun et al. 1996, Zych et al. 2007). Further ways of detectioning QGP include locating background quasars which are gravitationally lensed by foreground galaxies (with multiple images of the background quasar separated by a few arcseconds or less), searching for binary quasars producing projected QGP (false positives for binary quasars), with the background quasar probing the host galaxy of the foreground quasar ( e.g. Hennawi et al. 2006; Hennawi \& Prochaska 2007; Bowen et al. 2006; Myers et al. 2008), or examining discrepant redshifts from the same objects using different reduction codes (the discrepancies come from different ways of handling unidentified emission lines, for instance).  This paper uses a new technique by searching for narrow emission lines in quasar spectra that do not match the redshift of the broad, prominent quasar emisson lines (Quashnock et al. 2008, Noterdaeme et al. 2010, Borthakur et al. 2010, this study), then doing imaging and follow-up studies of the implied emission line galaxies.

After this work began, Noterdaeme et al. (2010) performed a search for [O III] $\lambda$5007 emission from foreground galaxies in spectra of QSOs. They published 46 such objects. We are undertaking to derive the properties of the QSOs and the galaxies in these types of pairs also, this paper being the first of four planned. 

This paper is organized as follows.  In section 2, the use of the Sloan Digital Sky Survey (SDSS) for finding emission objects in spectra of QSOs is discussed. In section 3, the results of our search of SDSS data release 5 (DR5) are described. The particular properties of 23 galaxies that lie directly on sightlines to quasars are elaborated. In section 4, we summarize our results and give our conclusions.

\section{Locating Emission Line Galaxies in SDSS Quasar Spectra}

In some cases, a quasar may lie directly behind a foreground galaxy. If some part of the galaxy is within $1.5\arcsec$ of the quasar, some fraction of the light from the galaxy falls into the $3\arcsec$-diameter fibers of the SDSS spectrograph.  Poorer seeing conditions could also cause light from a galaxy located at more than $1.5\arcsec$ from the QSO to fall within the fiber. 

The first such case,  SDSSJ104257.5+074850.5 ($z=2.66$, m$_{i}=19.03$), showing a strong radio source with a prominent foreground galaxy overlapping, was found while inspecting a large number of quasar spectra from the quasar absorption catalog of York et al. (in preparation;  see York et al. 2005, 2006; Vanden Berk et al. 2008; Khare et al. 2012).  Henceforth, quasars are referred to with the notation Q followed by the truncated four digit RA and declination, e.g., Q1042+0748 (epoch 2000). The clear detection of narrow emission at $z=0.033$ is seen in [O II] $\lambda3728.5$, H$\beta$ (4862.7), [O III] $\lambda\lambda4960.3, 5008.2$, H$\alpha$ (6564.6), and [S~II] $\lambda\lambda6718.3, 6732.7$. The lines of [N II] $\lambda\lambda6549.9, 6585.3$ are notably weak or missing. We use vacuum wavelengths, with close pairs at appropriate wavelength listed by Vanden Berk et al. (2001).  An exhaustive study of the foreground galaxy was undertaken by Borthakur et al. (2010). Figures 2 and 3 of Borthakur et al. (2010) show high resolution images of Q1042+0748. 

The redshift of Q1042+0748 is too high to expect the signature of the QSO host galaxy to show up in imaging. Therefore, the detected extended emission must be from a lower redshift intervening galaxy, most likely the galaxy corresponding to the z=0.033 emission lines. An effort was made to find other such images in the DR4 quasar catalog using an SQL search. The search asks for all objects that were targeted for spectroscopy as galaxies, based on the extended shape, but classified as quasars, based on the resulting spectrum. Twelve objects were found, including the prototype object Q1042+0748. No other galaxy foreground to a quasar image was found. A more sophisticated algorithm may, in the future, produce additional candidates, but we decided to try a different method. 

Any emission lines in the galaxy spectra show up in the quasar spectra as unidentified lines at a redshift different from the dominant quasar emission line redshift.  The unidentified lines are saved in the SDSS reduction files as significant emission lines with no identification. Our use of these data is described below. (In principle, the absorption spectra of an overlapping, foreground galaxy could also be isolated from the QSO spectrum (Vanden Berk et al. 2006)).  We therefore used a spectroscopic method of locating QGP, searching among these unidentified lines in the original DR5 spectroscopic reduction files. This method has the benefit of returning even small galaxies at small impact parameters, which could possibly be lost within the PSF of the QSO. The spectroscopic pipeline used to reduce quasar spectra uses an automated method to identify and classify the quasars, galaxies, and stars that were targeted for spectroscopy (Richards et al. 2002, Strauss et al. 2002, Beers et al. 2006). In the process of reducing spectra of targeted quasars, emission lines were found in some spectra with widths $>1000$ km s$^{-1}$. These formed consistent patterns in redshift that revealed the objects as quasars. Unidentified narrower lines were separately saved and can be searched for with an SQL search.

The SQL search requirements consisted of strong H$\alpha$ emission between 6565\AA ~and 9100\AA  (corresponding to a redshift range of $0 < z < 0.38$), FWHM of $<5$ pixels ($<560$ km s$^{-1}$), and a flux greater than $5 \times 10^{-17}$ ergs cm$^{-2}$ s$^{-1}$. The flux value was picked as a conservative value that would produce the clearest systems. The unidentified line strengths determined in the spectroscopic pipeline were used for the search. About 4000 indications were found (mostly false positives). The true detections were picked out by hand by confirming other emission lines present at the same redshift as the candidate H$\alpha$ line or by rejecting the supposed detection as a spectroscopic artifact. 

A Kolmogorov-Smirnov (KS) test  between our sample of QSOs and the Schneider et al. SDSS DR7 QSO catalog (Schneider et al. 2010) reveals a significance of 0.003\% corresponding to a maximum absolute discrepancy of 0.489 in r-band magnitude (using DR7 magnitudes for our sample). We have also run the KS test regarding QSO redshift between these two samples and found a significance of 39.84\% corresponding to a max absolute discrepancy of 0.19. This implies that our sample is significantly different than the overall QSO population found in the SDSS when looked at from the r-band. In subsequent papers our sample will be increased by a factor of three, making our sample of QSO magnitudes more robust. Noterdaeme et al. (2010) discuss the effect of QSO luminosity on detecting overlapping galactic emission lines. As one would expect, detections towards fainter QSOs are easier because lower signal-to-noise (S/N) ratio is necessary. However, S/N in SDSS spectra increases rapidly with QSO brightness, which allows for detections towards brighter QSOs. Very bright QSOs (m$_{i}<$17) are only a very small contribution to the statistics (Noterdaeme et al. 2010). Our measured reddening values support these results. Fainter QSOs tend to show smaller amounts of reddening, while the brighter QSOs found in our sample tend to show much more reddening, thus allowing them to be detected in our sample. We observe a slight difference in the redshift distribution that may be due to the effects of galactic emission lines being more difficult to detect on top of QSO emission lines. However, this difference has low significance ($\sim$60\%). Further effects biasing our sample include sky-line residuals and other strong emission line sources making it difficult to detect the galactic emission lines we seek.

Of these 4000, 23 were found to be positive detections.  False positives in this sample were primarily due to sky-line residuals, where a sky-line was mistakenly taken as H$\alpha$. Table~\ref{tbl-qgp} lists these detections and gives an index to each quasar (used in subsequent tables); the quasar name; the plate, fiber and mean Julian date (MJD) of observation that uniquely identifies the spectrum in the DR5 database; the quasar redshift from Schneider et al. (2007); the galaxy emission line redshift; and a quality grade indicating the  number of emission lines detected for each system. A grade of A means that four nebular lines were detected; grade B means that 3 nebular lines were detected; grade C means that 2 nebular lines were detected; and grade D means that only one line was detected. For a line to be counted, its flux density must be greater than $4 \times 10^{-17}$ ergs cm$^{-2}$ s$^{-1}$.  The [N II] and [S II] lines are counted individually, as are the two [O III] lines. Images of the 23 objects ($30\arcsec$ on a side) appear in Figure~\ref{fig-23} in order of increasing RA.  Numerical indices indicate order of increasing Plate number in SDSS. We note that H$\alpha$  for the system of Q0940+3415 falls outside the SDSS spectral range. However, given that this target was detected in the same search due to its strong [O III] and H$\beta$ presence, we have included it in this sample. We have not included it in figures or calculations involving H$\alpha$.

Noterdaeme et al. (2010) conducted a similar search for galaxies intervening with QSOs by locating the [O III] $\lambda\lambda$4959, 5007 doublet in the redshift range $0 < z < 0.7$. The overlap with our sample is in the range $0 < z < 0.38$. Their study found 46 systems by searching for [O III] while we found 23 systems by searching primarily for H$\alpha$. Of these, they detected 33 systems that we did not, while we found 10 systems that they did not. However, we could not have found 20 of these 33 objects, as H$\alpha$ falls outside our redshift range for these. Those systems found by Noterdaeme et al. contained in our sample are marked in Table~\ref{tbl-qgp}. All but two of our systems have detections in [O III]$\lambda$5007, and these two systems were also not detected by Noterdaeme et al. The 10 systems in our sample not found by Noterdaeme et al. were among the lowest in flux value for [O III] or had no detected [O III] at all, perhaps explaining why they were missed.

\section{Data Analysis}

\subsection{Deblending Quasar-Galaxy Pairs}

Unless the SDSS identified the quasar and the galaxy as two distinct objects (which only occurred for three systems), no attempt was made by the SDSS data pipeline to deblend the color data for the two objects.  The SDSS database deblended the three systems it identified as a QSO and a galaxy separately by masking one object when calculating the data for the other object.  In our attempt to deblend galaxy and quasar for the other 20 systems, we used Image Display Paradigm 3 (IDP3, an image manipulation program; Stobie 2006) to fit a point spread function (PSF) to the quasar image. The PSF was chosen as a star within the same field as the quasar, then matched to the magnitude of the quasar and subsequently subtracted, effectively removing the quasar. This allows us to perform photometry on the galaxy and quasar separately. 
 
The size of visible galaxies is recorded in arcseconds. The angular offset from the center of the galaxy to center of the quasar is recorded in arcseconds, from which the impact parameter is calculated and recorded in kpc. We do not find inconsistancies in impact parameter derived from different bands.  We assume a Hubble constant of 70 km s$^{-1}$ Mpc$^{-1}$, $\Omega_{m}=$ 0.3, and $\Omega_{\Lambda}=$0.7.  
 
Table ~\ref{tbl-offset} records our calculations, giving a running index (numbers increase with increasing plate number), the separation in right ascension and declination of the quasar from the galaxy center (in pixels, 0.396\arcsec each), the net offset in arcseconds, the corresponding impact parameter in kpc, the length and the width for the galaxy in arcseconds, and the position angle of the major axis (east from north).

\subsection{Broad-band flux measurements}

The galaxies found in each image are fitted with different models. The flux from a point source object, such as a star or a quasar, is best fit with point spread function model, however the flux from a galaxy is best fit with a matched galaxy model. The SDSS uses two models to fit galaxy flux profiles: a pure exponential and a deVaucoulers profile. For each object we have recorded the point spread function magnitude (psfMag), deVaucouler magnitude (devMag), and exponential magnitude (expMag) in the r-band as determined by SDSS pre-PSF subtraction.  We have also recorded the Petrosian radius in the r-band (petroRad), which is a measure of the size of the object. Even for galaxies that are not visible, the Petrosian radius and the model magnitude can indicate if the profile of the quasar deviates from that of a point source.  Quasars with intervening galaxies often have Petrosian radii larger than other point sources (stars) of the same field, and their model magnitudes commonly correspond to one of the galactic profiles rather than a PSF profile.  Particularly, the objects Q0239-0705, Q1610+5007, Q1143+5203, and Q1452+5443 (with indices 4, 9, 11, and 18) do not have a galaxy large enough to measure, but their Petrosian radii are larger than a usual point source (usually below 1.4) and their model magnitudes correspond to one of the galactic profiles. Table~\ref{tbl-phot} gives the three SDSS r magnitudes for different models of the light distribution in the initial SDSS reductions (Stoughton et al. 2002). The column marked ``dev'' is the De Vaucouleur magnitude, usually indicating a red, elliptical galaxy. The column marked ``exp'' uses the best-fit magnitude to an exponential light distribution in both axes, usually the signature of a spiral galaxy. The psf magnitude assumes a point source, where the psf profile is determined dynamically from stars in the near field.  Table~\ref{tbl-phot} lists these values and a superscript indicates which of the three models gave the best fit.

\subsection{Emission line measurements}

Found emission lines have been measured using the SPLOT package in IRAF.  The emission line flux density is measured in the standard SDSS units of $10^{-17}$ ergs cm$^{-2}$ s$^{-1}$.  All measurements of wavelengths and FWHM are in angstroms. Error values are found by measuring a featureless region of the spectrum near the emission line. Figure~\ref{fig-spectra-23} shows the spectra of each object from which the emission lines were measured and Table~\ref{tbl-emission} lists the measurements of all 9 emission lines for each of the 23 objects in our sample. 

\section{Results}

\subsection{Dust measurements along QSO sight-lines}

We proceed with photometric and spectral measurements of the galaxy and the quasar in each field. The magnitudes measured are inverse hyperbolic sine (asinh) magnitudes, described in detail in Lupton et al. (1999). The SDSS photometric data is taken under five filters in the imaging array (u, g, r, i, z) which have effective wavelengths of 3590~\AA, 4810~\AA, 6230~\AA, 7640~\AA, and 9060~\AA ~respectively (Fukugita et al. 1996, Gunn et al. 1998). The calculation of asinh magnitudes from total counts is performed as described at the SDSS photometric flux calibration webpage\footnote{http://www.sdss.org/dr5/algorithms/fluxcal.html\#counts2mag}.

The (g-i) color excess for the quasar is calculated from the de-convolved PSF magnitudes and compared with the median (g-i) value for quasars at the same redshift, taken from Richards et al. (2003). The difference between the measured (g-i) and the (g-i)$_{median}$ value from Richards et al. is recorded as $\Delta$(g-i).  This value is the observer-frame color excess and is a measure of the extinction of the quasar due to dust in the absorber.  

York et al. (2006) showed that quasars with absorption systems (QSOALS) were more likely to have larger positive values of $\Delta$(g-i), corresponding to a more reddened object.  The measured values of  $\Delta$(g-i) for quasars in our sample range between -0.255 and 0.650 magnitudes and show an even stronger preference for larger positive values than demonstrated in York et al. (2006).  The average value for all 23 found systems is 0.173 magnitudes.  A value $>0.2$ implies that the absorber galaxy has enough dust to significantly redden the quasar. Khare et al. (2012) find an average value for quasars in SDSS DR7 to be 0.16, with a median value of 0.085 at $z<2.8$.  Figure~\ref{fig-dust} shows the fraction of our sample per reddening bin, in addition to points relating the quasar redshift vs. $\Delta$(g-i). 

An important next step is to measure the extinction curves to determine their slope and the presence or absence of the 2175~\AA ~bump.  Most QSOALS have no bump and a steep UV rise shortward of 2000~\AA~ (York et al. 2006),  though recently bump detections have been made in several studies (e.g. Srianand et al. 2008, Eliasdottir et al. 2009, Jiang et al. 2010). These two parameters may distinguish low redshift galaxies from high redshift galaxies, or reveal whether those features  of the extinction curve are related to the mass and age of a galaxy.  The molecular content and the elemental content can also be measured to find the universality of the depletion patterns seen in the Milky Way and the Magellanic Clouds, thought to be indicative of  the presence and composition of dust grains. Expanded versions of the current sample will allow a statistical approach to determining the nature  of interstellar dust.

From the observer-frame color excess, we have also calculated the absorber rest-frame color excess, E(B$-$V)$_{(g-i)}$ as described in York et al. (2006).  This calculation uses the formula:

\begin{equation}
E(B-V)_{(g-i )} = \frac{\Delta(g-i)(1 + z_{abs})^{-1.2}}{1.506} 
\end{equation}
Here, z$_{abs}$ is the redshift of the absorber, in this sample, the galaxy producing emission lines.  The values found indicate significant reddening, with values as high as E(B-V)$_{(g-i)}= 0.37$, or A$_{v}= 1.11$ for an SMC extinction curve.  The most reddened objects are Q0013+0024 (number 3, A$_{v}=0.72$), Q1605+5107 (number 7, A$_{v}=0.57$), Q1610+5007 (number 9, A$_{v}=0.84$), Q1143+5203 (number 11, A$_{v}=1.11$), and Q1011+0619 (14, A$_{v}=1.02$). These values are all above the extinction in the Galaxy  and the LMC at which molecular hydrogen appears (Rachford et al. 2002, Tumlinson et al. 2002).  Such high values indicate we might expect these targets to be good candidates for studies of DIBs, interstellar molecules, and silicates. At higher redshifts, Srianand et al. (2008) report detections of dusty absorbers in 21-cm. Their two targets, at $z_{em}=1.89$ and 1.65, have A$_{v}=0.59$ and 0.75 respectively for an SMC dust model. It can be seen that even compared to these higher redshift QSOs, the values for our five most reddened objects are as high or higher. Kulkarni et al. (2011) also detect silicates in dusty absorbers with background QSOs in the redshift range $0.871<z_{em}<2.182$ with A$_{v}$ values in the range 0.89 - 1.71. These systems exhibit significantly more reddening than our sample with values comparable to Lyman break galaxies. This in combination with star formation rates (SFR; discussed below) suggests that systems studied in Kulkarni et al. (2011) are chemically more evolved than those in our current sample. Ellison et al. (2008) state that the best place to locate DIBs may be in those systems with high reddening, especially those with large Ca II and Mg II equivalent width (W). They have one detection of a DIB corresponding to an A$_{V}$ of 0.71. Table~\ref{tbl-phot} details our photometric data for the quasar and galaxy, as well as estimates of extinction for the quasar through the galaxy. The columns are the index; the SDSS i magnitude of the quasar; the (g-i) color index for the quasar; the SDSS reddening index, $\Delta$(g-i); the color excess E(B-V)$_{(g-i)}$ for the quasar; the r-band apparent magnitude of the galaxy; the (u-r) of the galaxy; the r-band absolute magnitude of the galaxy; and the luminosity in the r-band. 
 
Figure~\ref{fig-Ebcolor} shows E(B-V)$_{(g-i)}$ vs. b (kpc) and E(B-V)$_{(g-i)}$ vs. (u-r). E(B-V)$_{(g-i)}$ and b show a weak anticorrelation with a Spearman correlation coeffiecient of r$_{s}$=-0.53. Similarly, E(B-V)$_{(g-i)}$ and color are only very weakly correlated with r$_{s}$=0.25.

\subsection{Measurements from galactic emission lines}

For galaxies that have H$\alpha$ and [O II] lines  we can estimate the SFR of the galaxy using the relation from Kennicutt (1998):

\begin{equation}
SFR(M_{\sun} yr^{-1}) = 7.9\times10^{-42} L(H\alpha)(ergs ~s^{-1})
\end{equation}

\begin{equation}
SFR(M_{\sun} ~yr^{-1}) = 1.4\times10^{-41} L([O II])(ergs ~s^{-1})
\end{equation}
where L(H$\alpha$) is the luminosity of H$\alpha$ emission and L([O II]) is the luminosity of [O II] emission. Table~\ref{tbl-SFR} lists our spectroscopic calculations for the targets: the H$\alpha$/H$\beta$ ratio, the SFR calculated from H$\alpha$ and [O II] before extinction correction, and SFR from H$\alpha$ and [O II] corrected for extinction.  Note that all reported SFR are lower limits, as a large fraction of light from the galaxy falls outside the SDSS spectral fiber. Knowing the impact parameter and extent of the galaxy, we can correct for this fiber loss by estimating the fraction of the galaxy within 1.5\arcsec of the QSO and using it as a correction factor for the flux measured within the fiber. However, we do not expect the emission in any particular line to be uniform across the entire galaxy, so the best way to determine the total flux is follow up spectroscopy targeting the galaxies themselves. We find a range of corrections, going from no correction at all (entire visible galaxy falls within 1.5\arcsec; e.g. Q0239-0705 and Q0940+3415) to 0.81 (81\% of galaxy falls outside 1.5\arcsec; e.g. Q1501+5710). We find that the SFR for all systems are very low, ranging from 0.003 to 1.44 uncorrected for extinction in H$\alpha$. According to P\'eroux et al. (2011b), DLA and subDLA systems also have low SFR. Additionally, Figure 8 from Kulkarni et al. (2006) reports low SFR ($<$5 M$_{\sun}$ yr$^{-1}$) for DLA systems, suggesting our sample is consistent with the low global SFR in DLAs. Westra et al. (2010) determine the H$\alpha$ luminosity function in the redshift range 0.100 $<z<$ 0.377. Their Figure 4 plot the H$\alpha$ luminosity against redshift. Our L(H$\alpha$) values fall directly in the mid range in all cases, showing excellent agreement (see also Fujita et al. 2003 and Shioya et al. 2008). Additionally, we find good agreement between the H$\alpha$ SFR in our sample and that of the nearby field galaxies in Kewley et al. (2002). In comparing our SFR values to this sample, we find an average extinction-corrected SFR$_{H\alpha}$ of 3.8 compared to their sample average of 3.5. Similarly, the medians of these samples are 0.6 and 0.7 respectively.

The ratio of H$\alpha$ and H$\beta$ emission line fluxes is also a good measure of the presence of extinction for the part of the galaxy falling within the SDSS spectral fiber. It should be noted that the paths illuminated by the internal emission lines and by the QSO differ, so the actual extinction values need not be identical. Given the value of about 2.88 for H II region gas (Osterbrock 1989), any value of H$\alpha$/H$\beta$ other than this would be an indication of extinction with case-B recombination. Figure~\ref{fig-hatohb} shows the distribution of extinction values, using H$\alpha$/H$\beta$. Overplotted are points for E(B-V)$_{(g-i)}$ vs. H$\alpha$/H$\beta$. We find only a weak correlation between the two with r$_{s}$=0.34. The reddening can be calculated as for the SMC extinction curve:

\begin{equation}
 E(B-V)_{H\alpha/H\beta} = \frac{1.086}{k(H\beta)-k(H\alpha)} ln(\frac{H\alpha}{2.88H\beta})
\end{equation}
with k(H$\alpha$) and k(H$\beta$) taken from Pei (1992). These results can be found in Table~\ref{tbl-SFR}. Figure~\ref{fig-ebvha_vs_ebvgi} compares the color excess of the quasar to that of the galaxy utilizing our two equations for E(B-V).  With a few exceptions, our sample indicates the gas directly in front of the quasar in our image is less dusty than the gas within the galaxy probed by the 1.5\arcsec radius spectral fiber, as show by Figure~\ref{fig-ebvha_vs_ebvgi}. This is consistent with our conclusion that these galaxies are largely late-type, and therefore dusty. We have little reason to expect the two quantities to be correlated, however, as they potentially sample different sight lines through the galaxy.  It should be noted that dusty lines of sight may be missed in our sample due to the QSO being too reddened. E(B-V)$_{H\alpha/H\beta}$ is measured from any part of the galaxy that falls within 1.5\arcsec of the quasar (SDSS spectral fiber diameter is 3\arcsec), while E(B-V)$_{(g-i)}$ is measured directly in front of the quasar. The dotted line on the graph marks the 1:1 boundary between the quantities.  It is possible the lower amount of dust along the line of sight to the QSO is related to impact parameter. We address this by plotting $\Delta$E(B-V) ($=$E(B-V)$_{(g-i)}$ - E(B-V)$_{H\alpha/H\beta}$) vs. b (kpc) in Figure~\ref{fig-ebvvsb} and find no significant relationship between the two quantities.

\section{Discussion}

\subsection{Galaxy colors}


When the pairs can be separated, the galaxies are blue with (u-r) $< 2$. Figure 8 from Strateva et al. (2001) depicts (u-r) distributions for 500 galaxies, spectroscopically determined to be either early- or late-type. The solid line indicates late-type galaxies and the dotted line indicates early-type galaxies. The mean (u-r) color for 500 well determined galaxies is $1.45\pm0.25$, at the center of the range of late type galaxies (Strateva et al. 2001). In Figure~\ref{fig-color}, we have plotted an analogous histogram to that of Strateva et al. for a subsample of 14 galaxies, that is, those of our 23 galaxies with (u-r) calculations, shown with the black line. While our sample is too small to show the peaks or color separation in late- and early-type galaxies, we can compare the range of our plot to that of Strateva et al. We can clearly see that all of our bins fall well within the range of late-type galaxies. The median value for the black histogram is (u-r)=1.43. The red histogram shows the (u-r) distribution from the SDSS magnitudes for the same 14 galaxies prior to deblending the galaxy and the quasar. We can see that prior to the deblending, the galaxies are systematically much bluer, with a median value of (u-r)=0.93. Most of these values fall blueward of even the late-type range in the Strateva et al. (2001) plot. This is due to the fact that the light from the QSO skews the magnitude values towards the bluer filters, causing them to be much brighter than if the galaxy were measured alone.  This result isn't so surprising, as we expect galaxies selected via emission lines will naturally tend towards late-types.

The E(B-V)$_{(g-i)}$ values found in our study indicate that many of the found objects are significantly reddened with an average E(B$-$V)$_{(g-i)}$ value of 0.101.  Thus, the average extinction in the observer frame and the absorber rest frame is significantly greater than the average extinction of non-absorber quasars as well as QSOALSs as measured in York et al. (2006). Figure~\ref{fig-extinction} below shows an analogous graph to Figure 5 from York et al. for our data.  A Kolmogorov-Smirnov test between the sample in this study and the much larger sample found in York et al. (2006) shows that the maximum absolute discrepancy between the two sets is 0.476, with a significance of 7.9$\times 10^{-3}$\%. The two samples compared in the test, those of this study and those of York et al., contained 23 and 809 data points respectively. The results indicate that our sample is significantly different than the sample in York et al., with our absorbers being more reddened. Again, Figure~\ref{fig-color} indicates that our absorber galaxies are primarily late-type, which are known to contain large amounts of dust compared to early-type galaxies, thus causing increased reddening in the background quasars.

Comparisons can also be made with systems with diffuse interstellar bands (DIBs), Ca II systems, and galaxies known to contain DLAs. Ellison et al. (2008) search nine Ca II-selected absorbers in the redshift range 0.07 $<z<0.55$ for DIBs. They report one detection with E(B-V)$=0.23$,  which is higher than typical values for Ca II selected systems and is generally higher than most of our sample but is comparable to our five most reddened systems. Wild et al. (2006) provide a sample of 37 Ca II absorbers and compare the degree of reddening in background QSOs. They find E(B-V) values in the range -0.008 to 0.417, with five out of six systems in their DR4 sample having E(B-V)$<0.2$. This is consistent with the values we have found. Of our most reddened objects, four of five have Ca II absorption. Regarding DLAs, Khare et al. (2012) report that only 10\% or less of DLAs in SDSS DR7 cause significant reddening in background QSOs. However, higher dust content in DLAs is found in those not selected on the basis of color. As such, dusty QSOs may be missed due to this color selection bias.  

Three of the pairs in our sample are photometrically separated in SDSS: QJ0943-0043 (object 2), QJ1605+5107 (object 7), and QJ1401+4141 (object 20). That is, SDSS recognizes the galaxy and the quasar as two separate objects, and performs photometric calculations for them separately. Table~\ref{tbl-deblended} shows the calculated values for our study and SDSS side-by-side for comparison. We find the (u-r) values to match within (u-r) of 0.02 for all three objects.  

\subsection{Absorption lines}

Equivalent widths or limits for interstellar Ca II H and K and Na I D1 and D2 are given in Table~\ref{tbl-ew}. The columns give the index, the wavelength and equivalent widths for the two Ca II lines (K and H), and the same for the two Na I lines (D2 and D1). The rest wavelengths are also listed in Table~\ref{tbl-ew}.  Figure~\ref{fig-cana} shows the relationship of Na I equivalent widths to Ca II equivalent widths. We find no correlation with r$_{s}$=-0.04. This should be taken with caution, however, given we only have a sample of 9 systems for this relationship.  There are two possible outliers, the point in the far upper left and the point in the far lower right. These points correspond to the galaxies towards Q0016+1356 and Q1238+1056 (objects numbered 10 and 19 respectively).  These objects have H$\alpha$ SFR corresponding to 0.40 M$_{\sun}$ yr$^{-1}$ and 0.35 M$_{\sun}$ yr$^{-1}$ respectively. Object 10 has a velocity spread of at least 300 km/s according to the equivalent width of Mg II. Object 19 could be a good candidate for a bunch of disk clouds with Ca II depletion.  

The relationships of W(Ca II) and W(Na I) absorption to impact parameter and extinction are given in Figures~\ref{fig-cana_b} and~\ref{fig-cana_ebv}.  We find no significant correlation between W(Ca II) and b or W(Na I) and b for $\alpha=0.05$. 
However, W(Ca II) and E(B-V)$_{(g-i)}$ are moderately anticorrelated with r$_{s}$=-0.77 ($\alpha=0.05$).  This is perhaps due to the depleted nature of Ca II, whereby dustier systems have narrower W(Ca II) because Ca II condenses out of the gas phase more readily. W(Na I) and E(B-V)$_{(g-i)}$ are not significantly correlated with r$_{s}$=0.14. Again, these samples are small, with only 8 and 9 data points for the figures respectively, decreasing our chances of finding a significant correlation.

\section{Summary and Conclusions}

We have measured the properties of the quasars and foreground galaxies in 23 fields containing quasar-galaxy pairs from the SDSS. These galaxies in front of quasars were detected via their narrow emission lines in the more dominant quasar spectra. The search detected H$\alpha$ emission, and the results were in turn inspected by eye for further emission lines and any visible galaxy counterpart in the SDSS imaging. By performing PSF subtraction to remove the quasar from the image, we were able to make measurements on the galaxy without light contamination from the very nearby quasar. From this, we were able to determine apparent and absolute magnitudes and reddening and extinction estimates. Utilizing the emission lines found in the spectra, we were also able to determine SFR and reddening estimates. 

We find no trends relating E(B-V)$_{(g-i)}$, \textit{b}, and (u-r) for the galaxies. Similarly, we find no trend relating Ca II or Na I absorption strength to \textit{b} or E(B-V)$_{(g-i)}$, nor between Ca II and Na I, given the few data points currently in our sample. 

We note that high resolution ground based observations of Ca II, CH, CH+, Ca I, Na I, and K I can be used to explore the properties of the gas in a larger array of galaxies than has been done heretofore. As radio telescopes improve in sensitivity, 21 cm absorption can be a measure of the generally weak radio sources that will be typical of this sample (Gupta et al. 2010). The diffuse interstellar bands can be observed and correlated with other gas properties. Ultraviolet observations can yield detailed extinction curves and abundances of the elements for comparison with high redshift QSO absorbers. 

Our sample size is still very small, but we have already begun the process of enlarging it by a factor of more than 4. Similar searches of SDSS have returned a further $\sim$100 pairs, which will be published in subsequent papers in this series. In this work we have begun comparing our small sample to color and extinction studies done by Strateva et al. (2001) and York et al. (2006). Our comparison is preliminary and will be much improved with the increased sample size, but so far our results fall within the ranges of the study by Strateva et al. (2001), with our sample being more reddened than the sample of York et al. (2006). The sample discussed in this work is yet too small to make any assumptions about the reddening ranges, however, the beginning color trend and results of the Kolmogorov-Smirnov test indicate they could be dusty, characteristic of disk-dominated galaxies. 

This is the first of a series of papers studying selection techniques and properties of galaxies intervening with background QSOs in the SDSS at low redshifts. The second paper of this series will address a selection technique very similar to that presented here, but by searching for nine galactic emission lines (including H$\alpha$). This will increase our sample size to a total of 50 targets. A third paper in this series will add a further $\sim40$ targets selected by [O III] emission as in Noterdaeme et al. (2010). A detailed comparison of the selection techniques will be addressed.

\section*{Acknowledgements}

We thank Michael Strauss for his early discussion and encouragement on QGP, and Joe Meiring for his continued discussions on the subject. LAS and VPK acknowledge partial funding from NSF grant AST/0908890 and from the NASA South Carolina Space Grant Consortium. We would also like to thank the anonymous referee for numerous constructive comments which improved this paper.

\pagebreak

\clearpage







\begin{figure}
\begin{minipage}{1.0\linewidth}
\begin{center}
\subfigure{
\includegraphics[trim = 0mm 0mm 0mm 0mm, clip, width=0.25\textwidth]{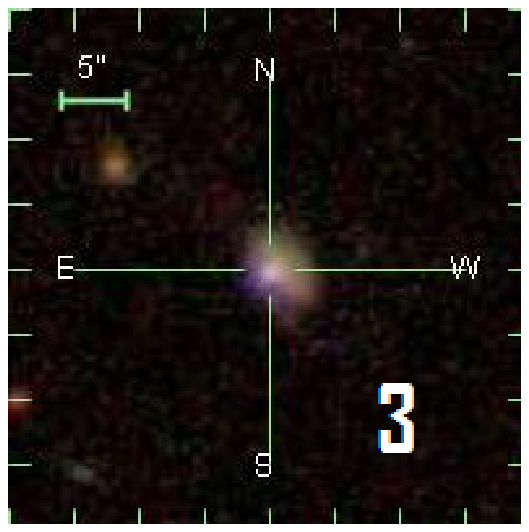}
}
\subfigure{
\includegraphics[trim = 0mm 0mm 0mm 0mm, clip, width=0.25\textwidth]{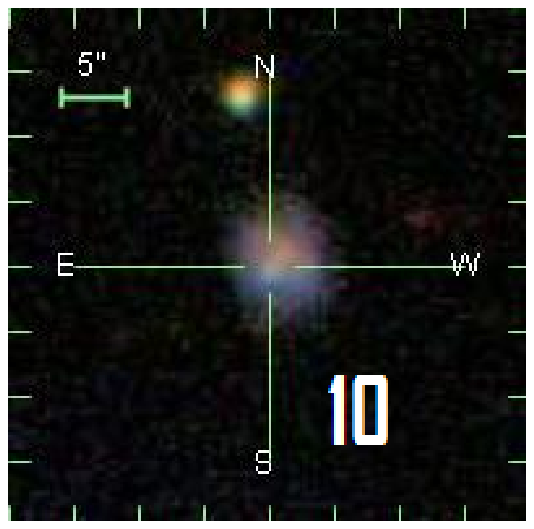}
}
\subfigure{
\includegraphics[trim = 0mm 0mm 0mm 0mm, clip, width=0.25\textwidth]{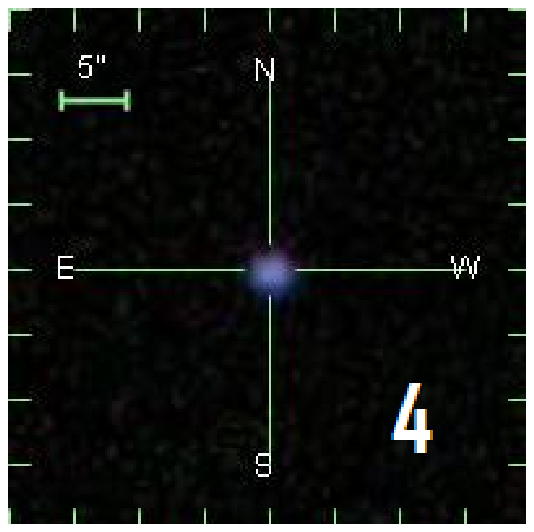}
}
\subfigure{
\includegraphics[trim = 0mm 0mm 0mm 0mm, clip, width=0.25\textwidth]{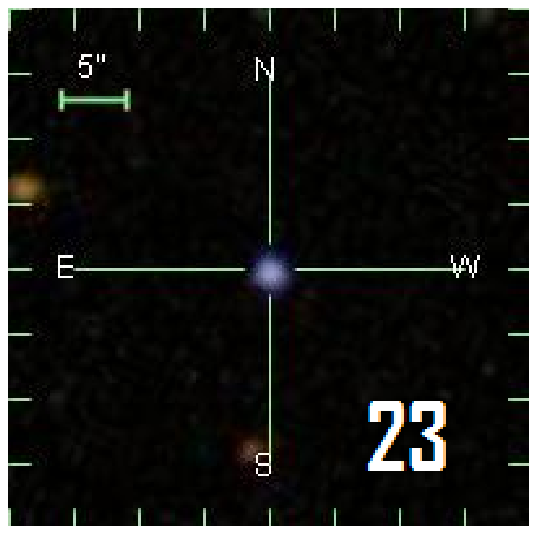}
}
\subfigure{
\includegraphics[trim = 0mm 0mm 0mm 0mm, clip, width=0.25\textwidth]{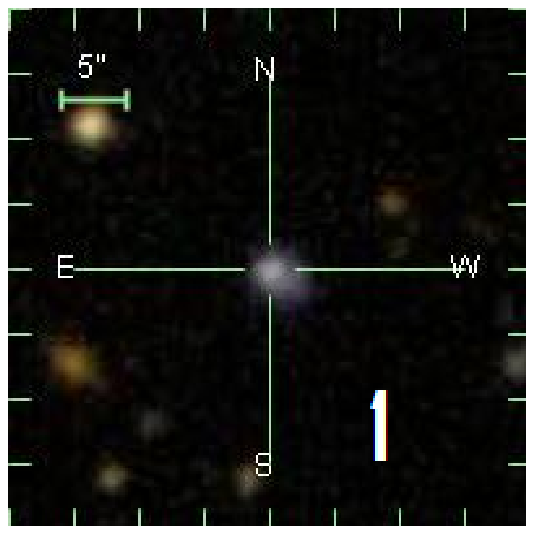}
}
\subfigure{
\includegraphics[trim = 0mm 0mm 0mm 0mm, clip, width=0.25\textwidth]{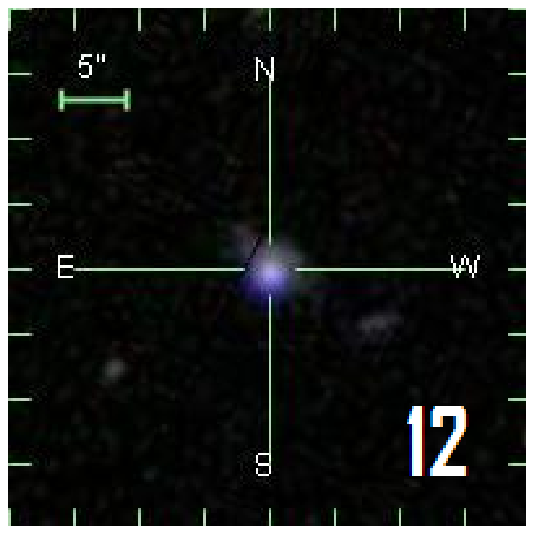}
}
\subfigure{
\includegraphics[trim = 0mm 0mm 0mm 0mm, clip, width=0.25\textwidth]{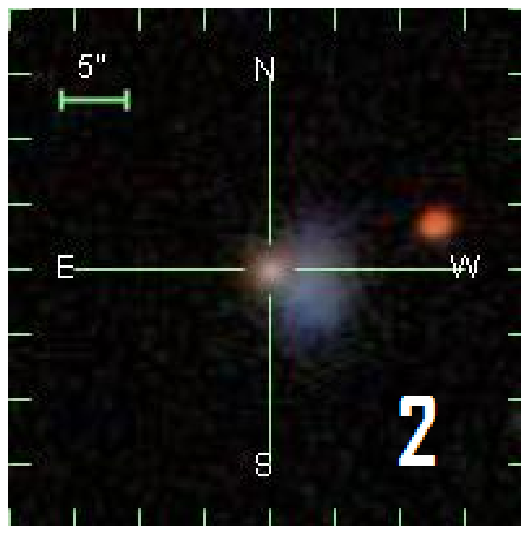}
}
\subfigure{
\includegraphics[trim = 0mm 0mm 0mm 0mm, clip, width=0.25\textwidth]{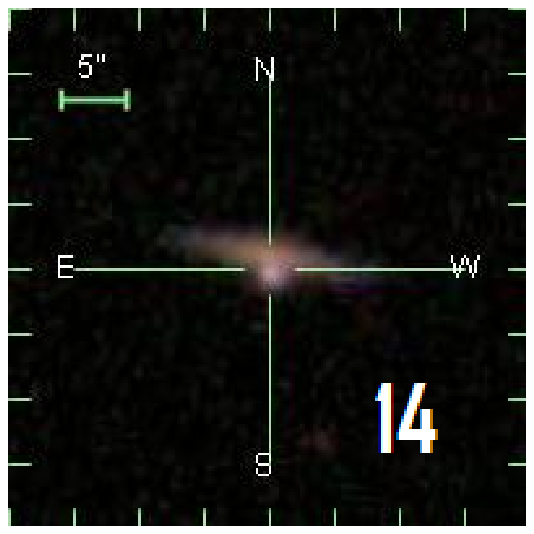}
}
\subfigure{
\includegraphics[trim = 0mm 0mm 0mm 0mm, clip, width=0.25\textwidth]{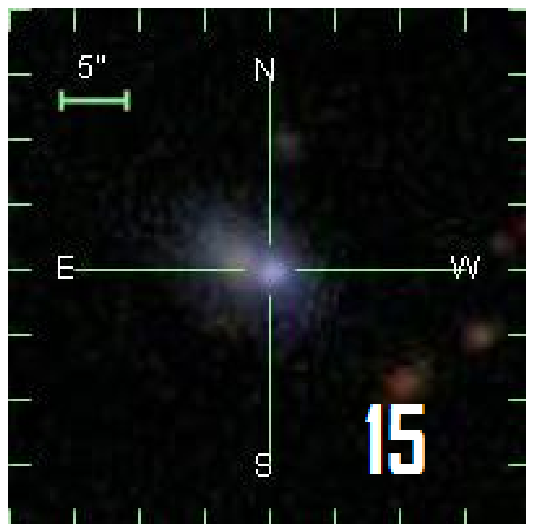}
}
\subfigure{
\includegraphics[trim = 0mm 0mm 0mm 0mm, clip, width=0.25\textwidth]{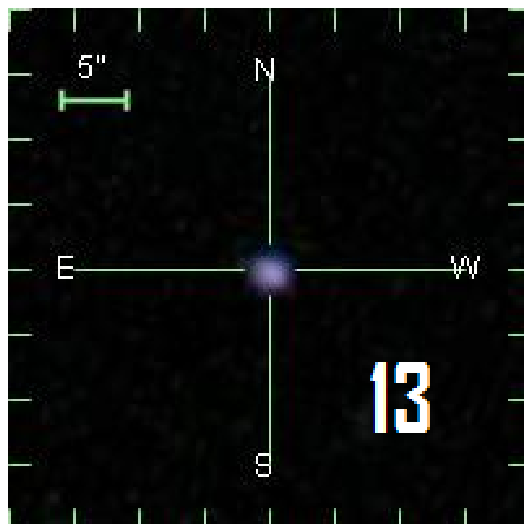}
}
\subfigure{
\includegraphics[trim = 0mm 0mm 0mm 0mm, clip, width=0.25\textwidth]{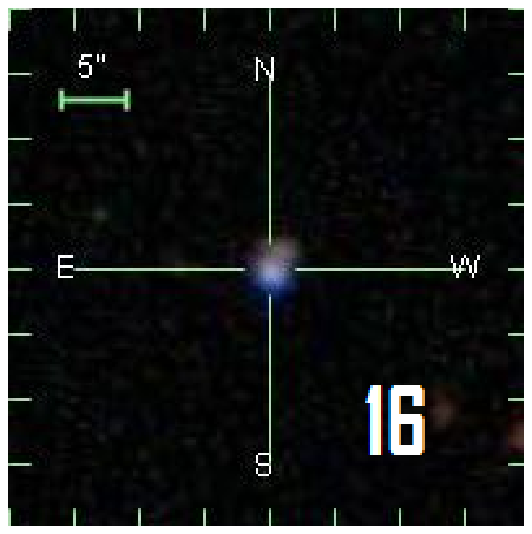}
}
\subfigure{
\includegraphics[trim = 0mm 0mm 0mm 0mm, clip, width=0.25\textwidth]{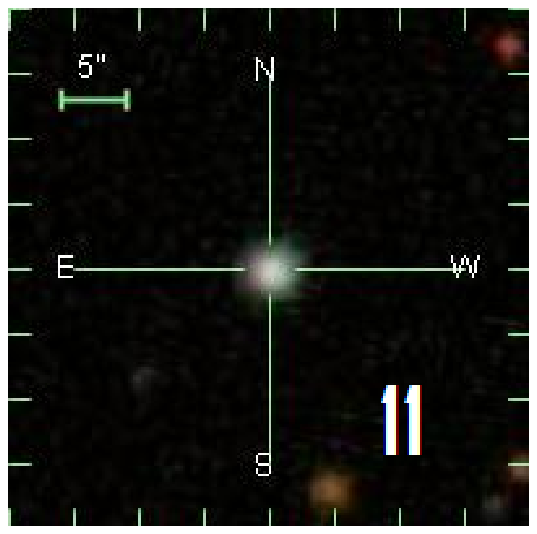}
}
\end{center}
\captcont{SDSS multicolor images of the 23 fields with a quasar intercepting a low-z galaxy, in order of increasing RA. Numerical values indicate order of increasing plate number. The scale of the images is indicated in the upper left hand corner. The orientation is north up and east left. }\label{fig-23}
\end{minipage}
\end{figure}

\begin{figure}
\begin{minipage}{1.0\linewidth}
\begin{center}

\subfigure{
\includegraphics[trim = 0mm 0mm 0mm 0mm, clip, width=0.25\textwidth]{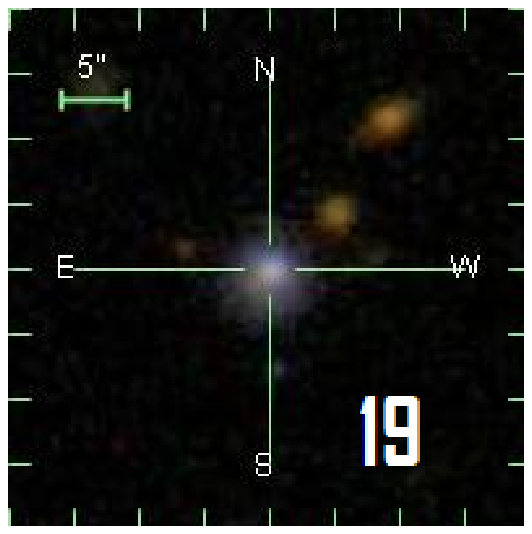}
}
\subfigure{
\includegraphics[trim = 0mm 0mm 0mm 0mm, clip, width=0.25\textwidth]{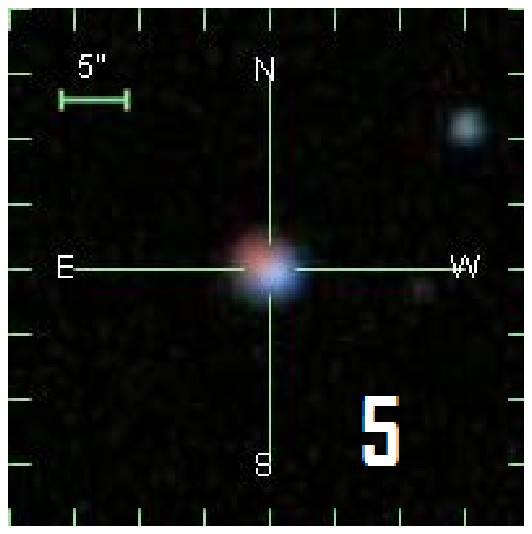}
}
\subfigure{
\includegraphics[trim = 0mm 0mm 0mm 0mm, clip, width=0.25\textwidth]{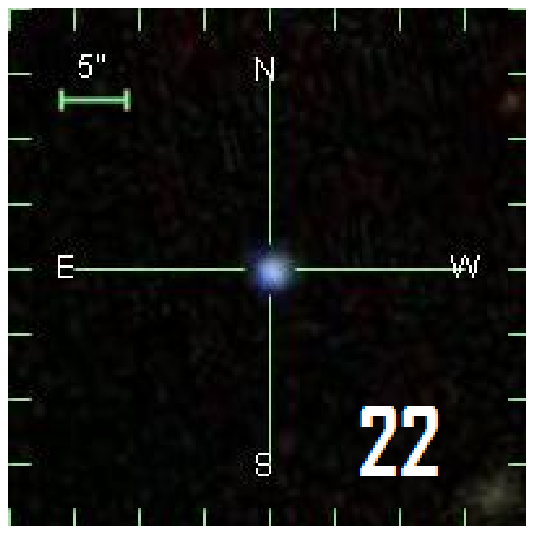}
}
\subfigure{
\includegraphics[trim = 0mm 0mm 0mm 0mm, clip, width=0.25\textwidth]{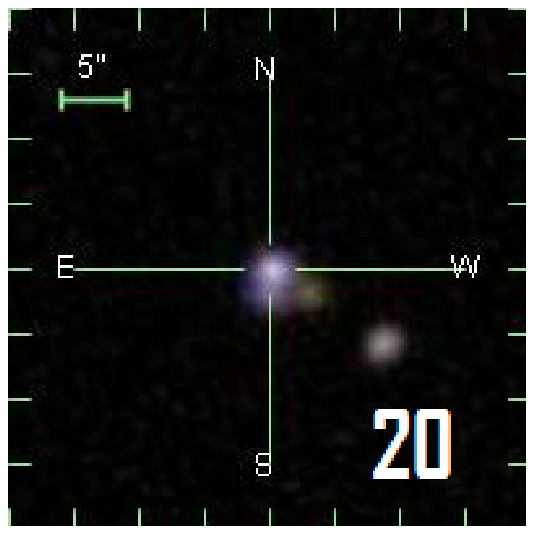}
}
\subfigure{
\includegraphics[trim = 0mm 0mm 0mm 0mm, clip, width=0.25\textwidth]{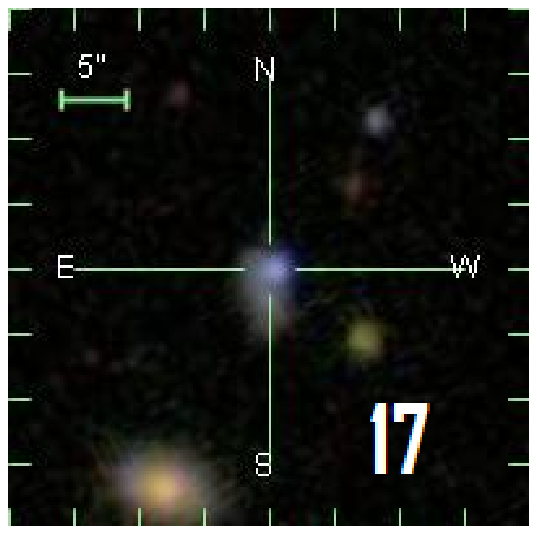}
}
\subfigure{
\includegraphics[trim = 0mm 0mm 0mm 0mm, clip, width=0.25\textwidth]{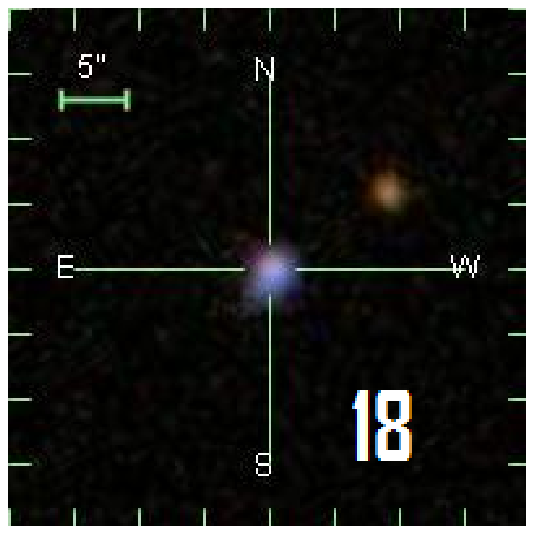}
}
\subfigure{
\includegraphics[trim = 0mm 0mm 0mm 0mm, clip, width=0.25\textwidth]{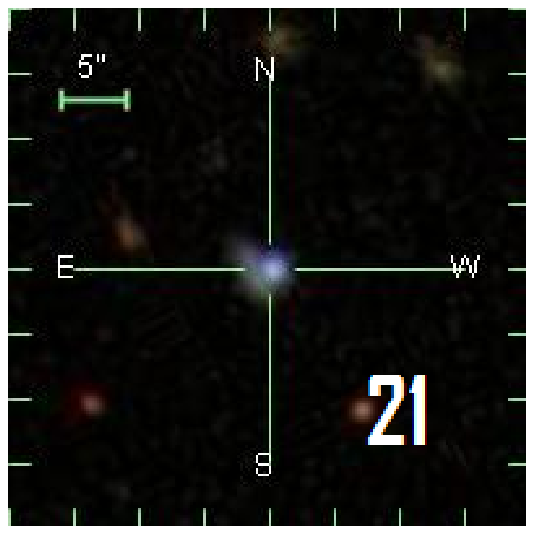}
}
\subfigure{
\includegraphics[trim = 0mm 0mm 0mm 0mm, clip, width=0.25\textwidth]{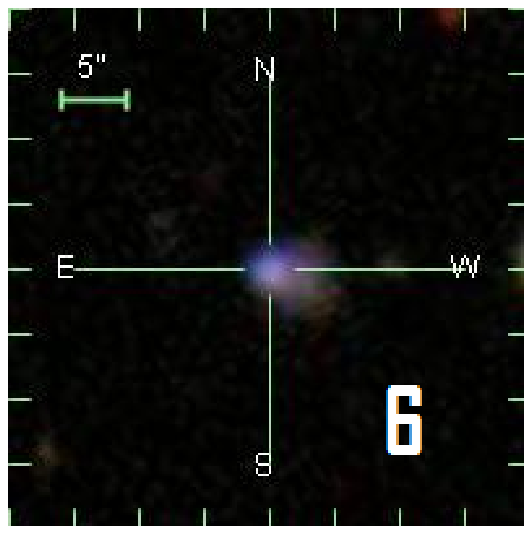}
}
\subfigure{
\includegraphics[trim = 0mm 0mm 0mm 0mm, clip, width=0.25\textwidth]{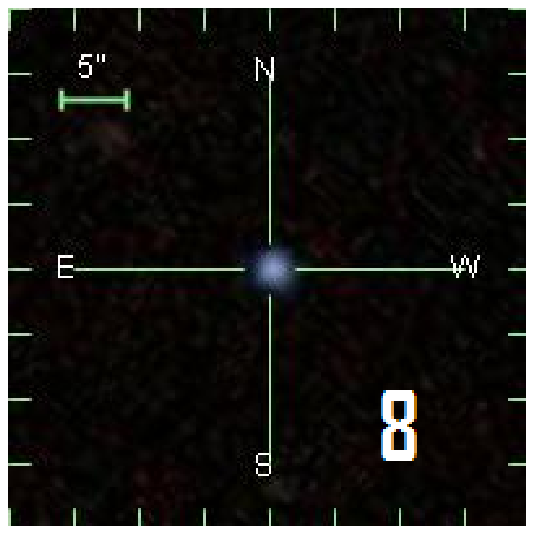}
}
\subfigure{
\includegraphics[trim = 0mm 0mm 0mm 0mm, clip, width=0.25\textwidth]{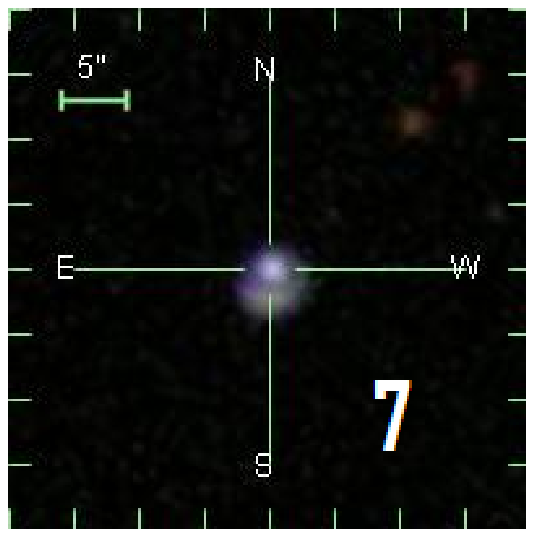}
}
\subfigure{
\includegraphics[trim = 0mm 0mm 0mm 0mm, clip, width=0.25\textwidth]{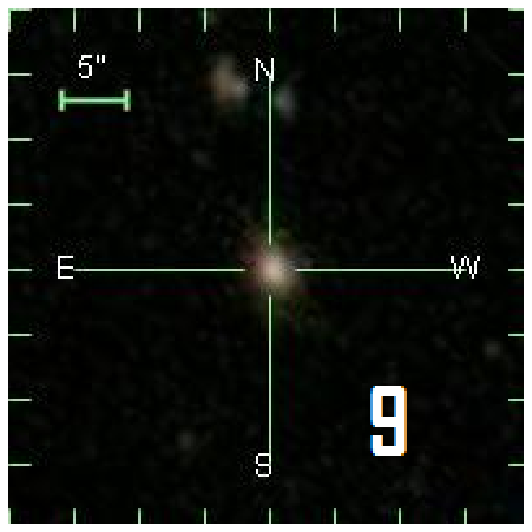}
}
\end{center}
\caption*{Continued.}
\end{minipage}
\end{figure}


\begin{figure}
\begin{minipage}{1.0\linewidth}
\begin{center}


\includegraphics[trim = 0mm 0mm 0mm 0mm, clip, width=0.65\textwidth]{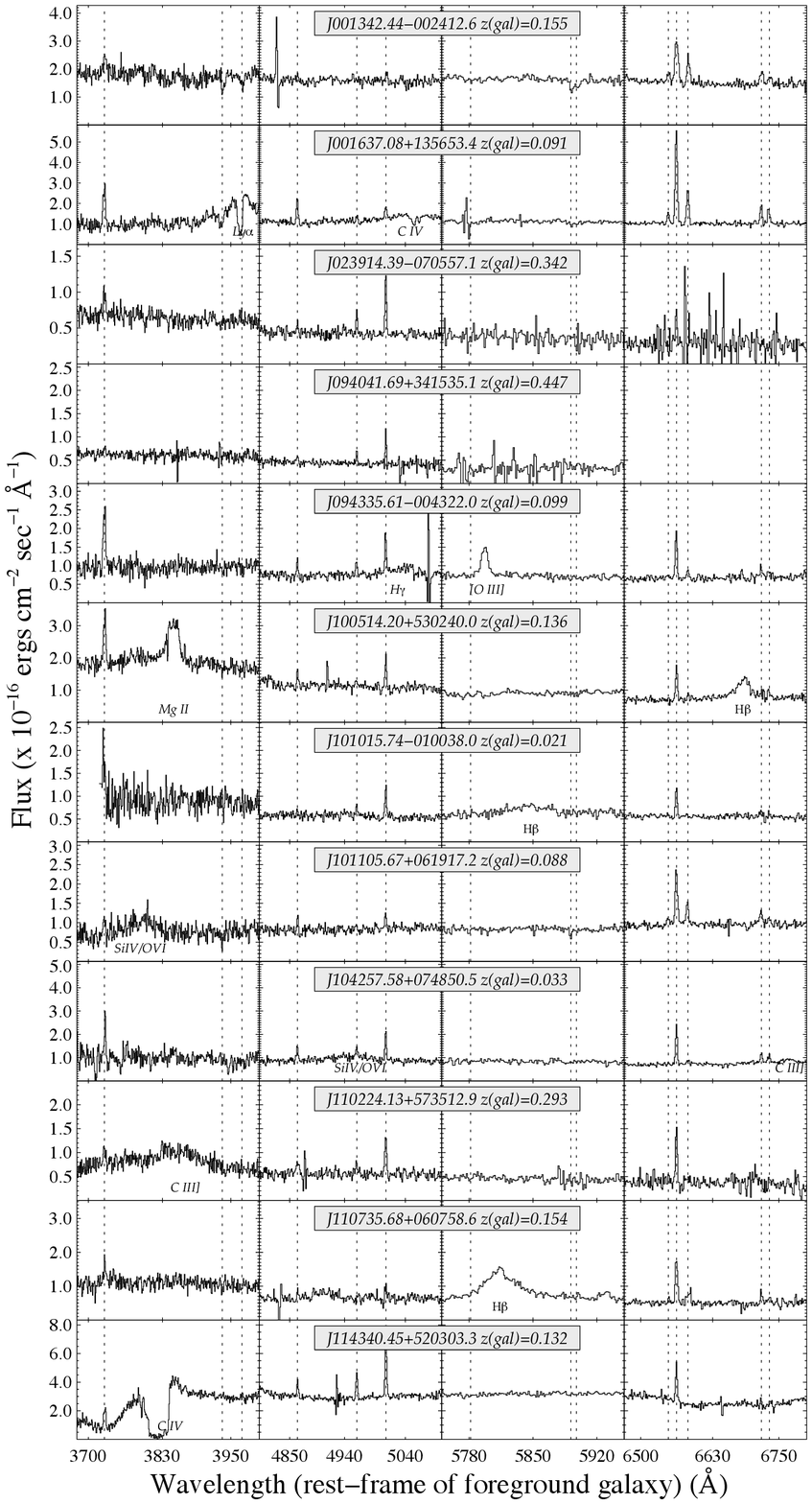}

\end{center}
\captcont{\tiny SDSS spectra for each of the 23 quasar systems. Selected regions of the spectra are shown. The dotted vertical lines mark the locations of (vacuum wavelengths) [O II] $\lambda$3728.38; Ca II absorption (3934.78~\AA, 3969.59~\AA); H$\beta$; [O III] $\lambda\lambda$4960.30, 5008.24; the diffuse interstellar band, 5782.10~\AA; Na I absorption (5891.58~\AA, 5897.56~\AA); [N II] $\lambda$6549.85; H$\alpha$; [N II] $\lambda$6585.28; and [S II] $\lambda\lambda$6718.29, 6732.67. Identifications of QSO emission lines are written onto the figures. Unmarked emission features, usually accompanied by adjacent absorption, are artifacts.}
\end{minipage}
\end{figure}

\begin{figure}
\begin{minipage}{1.0\linewidth}
\begin{center}

\includegraphics[trim = 0mm 0mm 0mm 0mm, clip, width=0.65\textwidth]{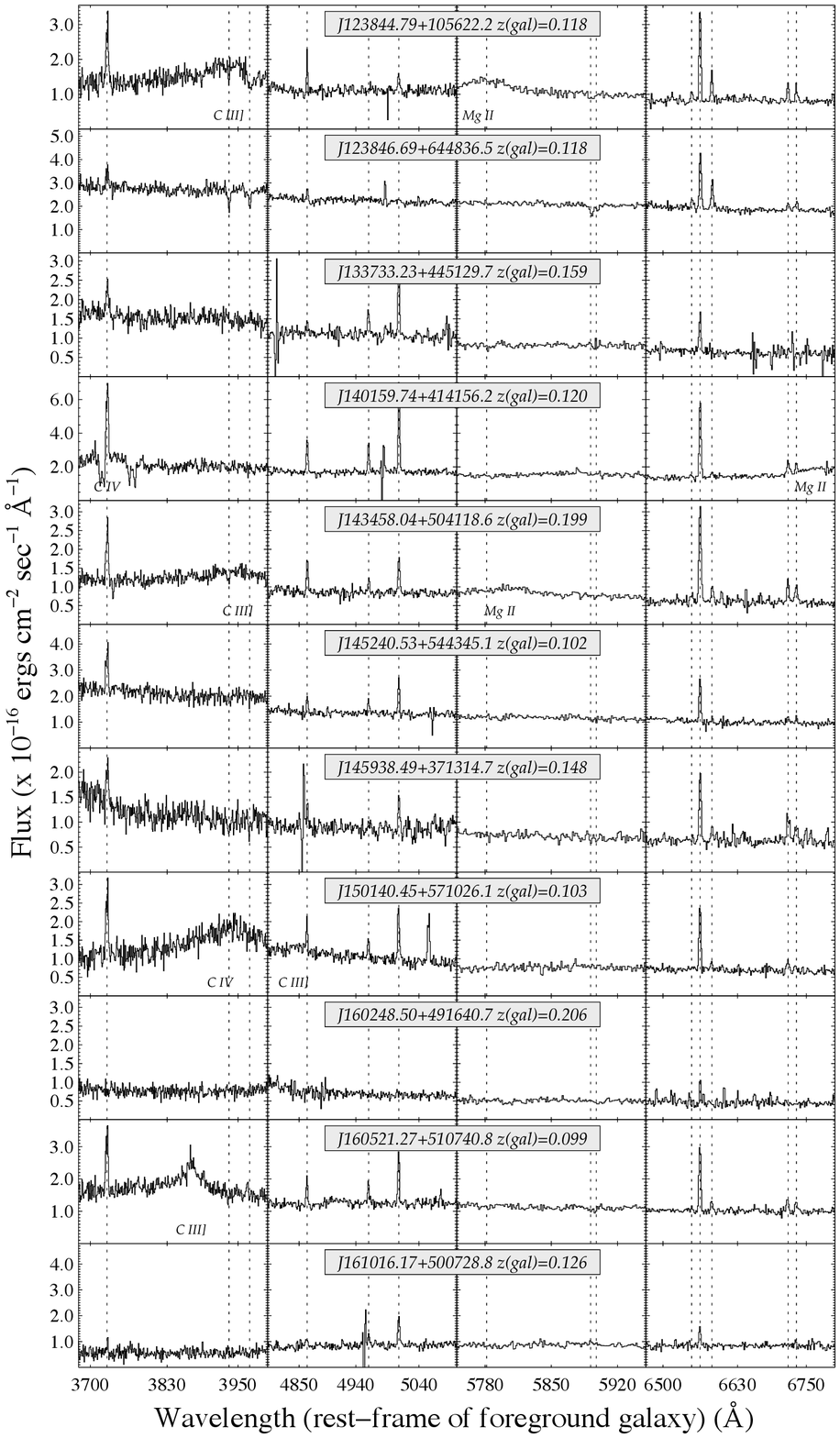}

\end{center}
\caption*{Continued.}\label{fig-spectra-23}
\end{minipage}
\end{figure}


\begin{figure}
\begin{minipage}{1.0\linewidth}
\begin{center}

\includegraphics[trim = 0mm 0mm 0mm 0mm, clip, width=0.9\textwidth]{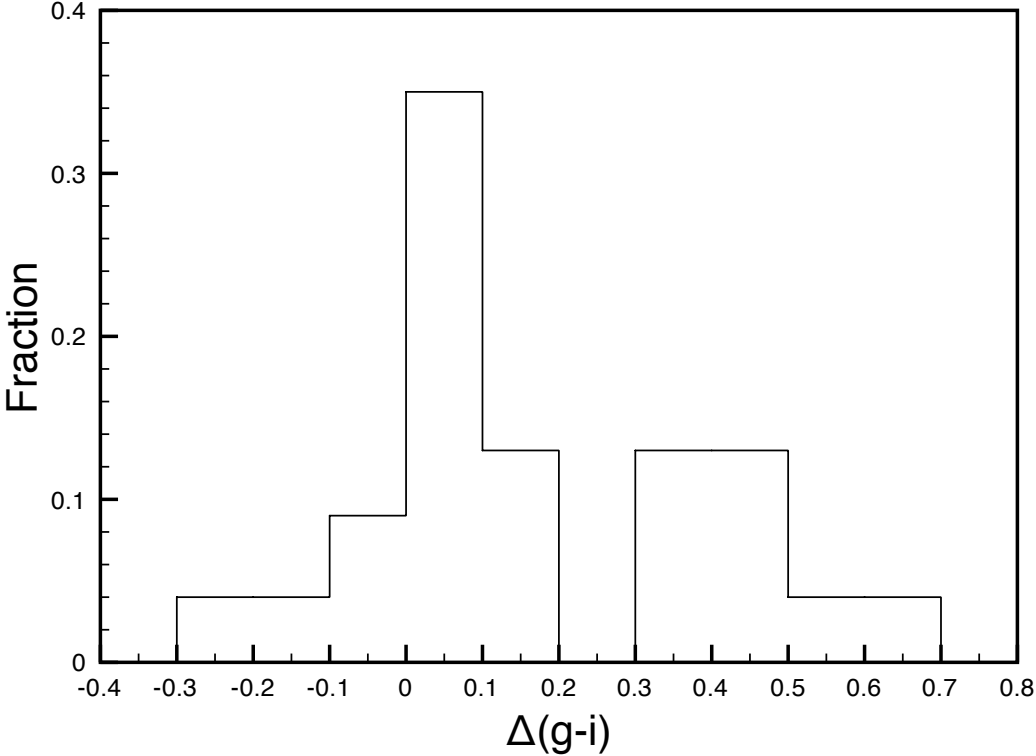}

\end{center}
\caption{Histogram showing the number of targets per $\Delta$(g-i) bin. }\label{fig-dust}
\end{minipage}
\end{figure}


\begin{figure}
\begin{minipage}{1.0\linewidth}
\begin{center}

\subfigure[]
{\label{fig-Ebcolor}
\includegraphics[trim = 0mm 0mm 0mm 0mm, clip, width=0.45\textwidth]{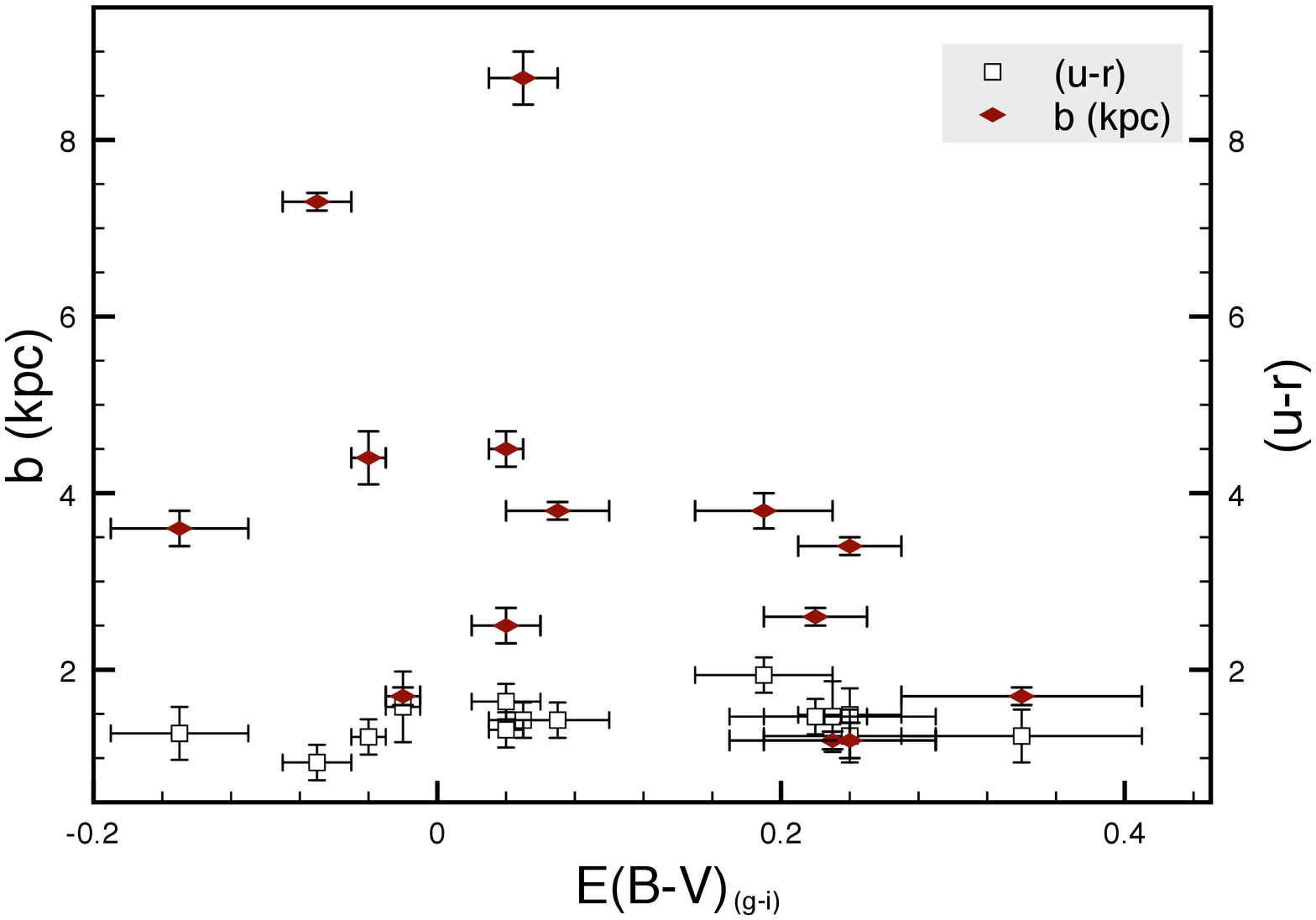}
}
\subfigure[]
{\label{fig-hatohb}
\includegraphics[trim = 0mm 0mm 0mm 0mm, clip, width=0.45\textwidth]{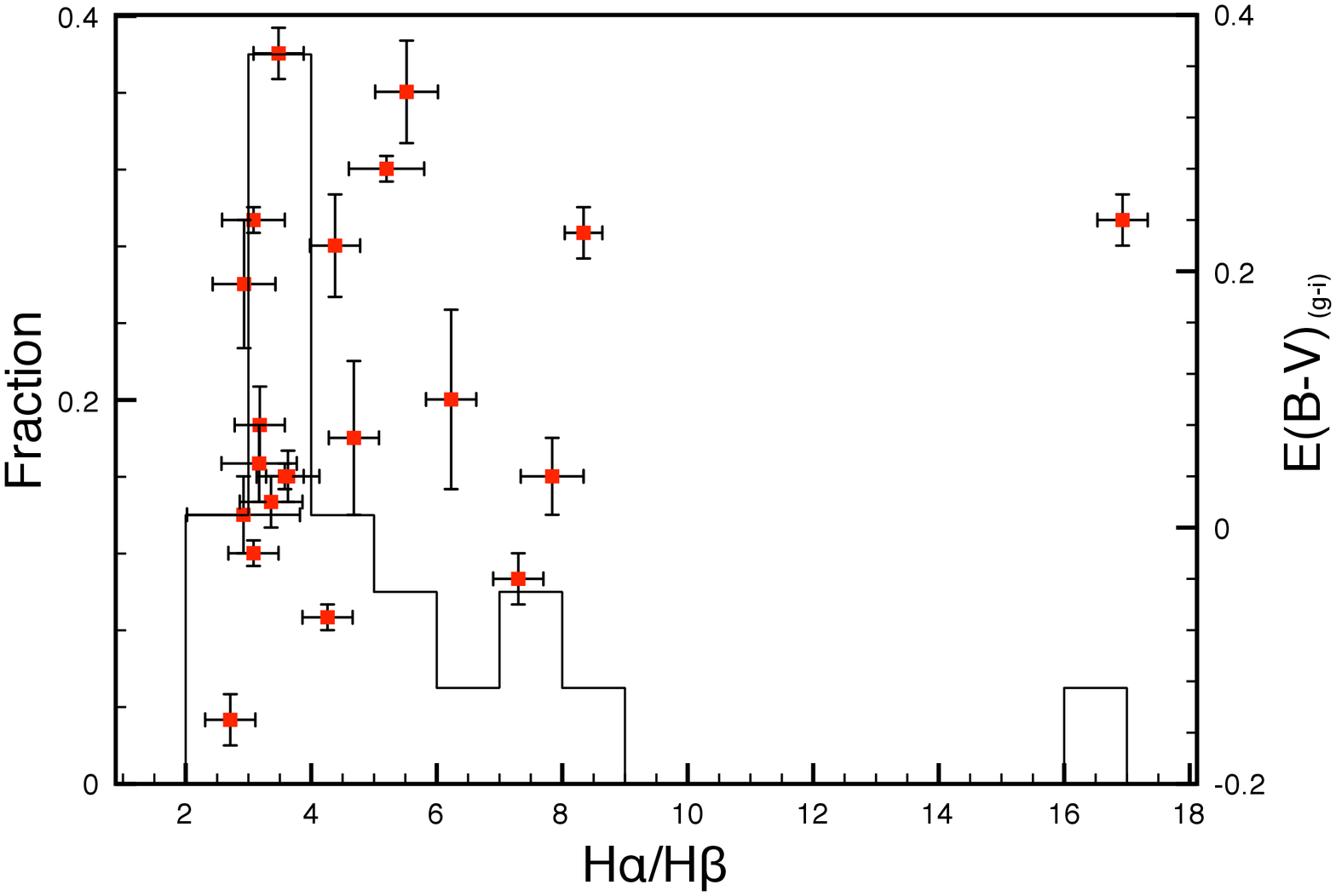}
}
\subfigure[]
{\label{fig-ebvha_vs_ebvgi}
\includegraphics[trim = 0mm 0mm 0mm 0mm, clip, width=0.45\textwidth]{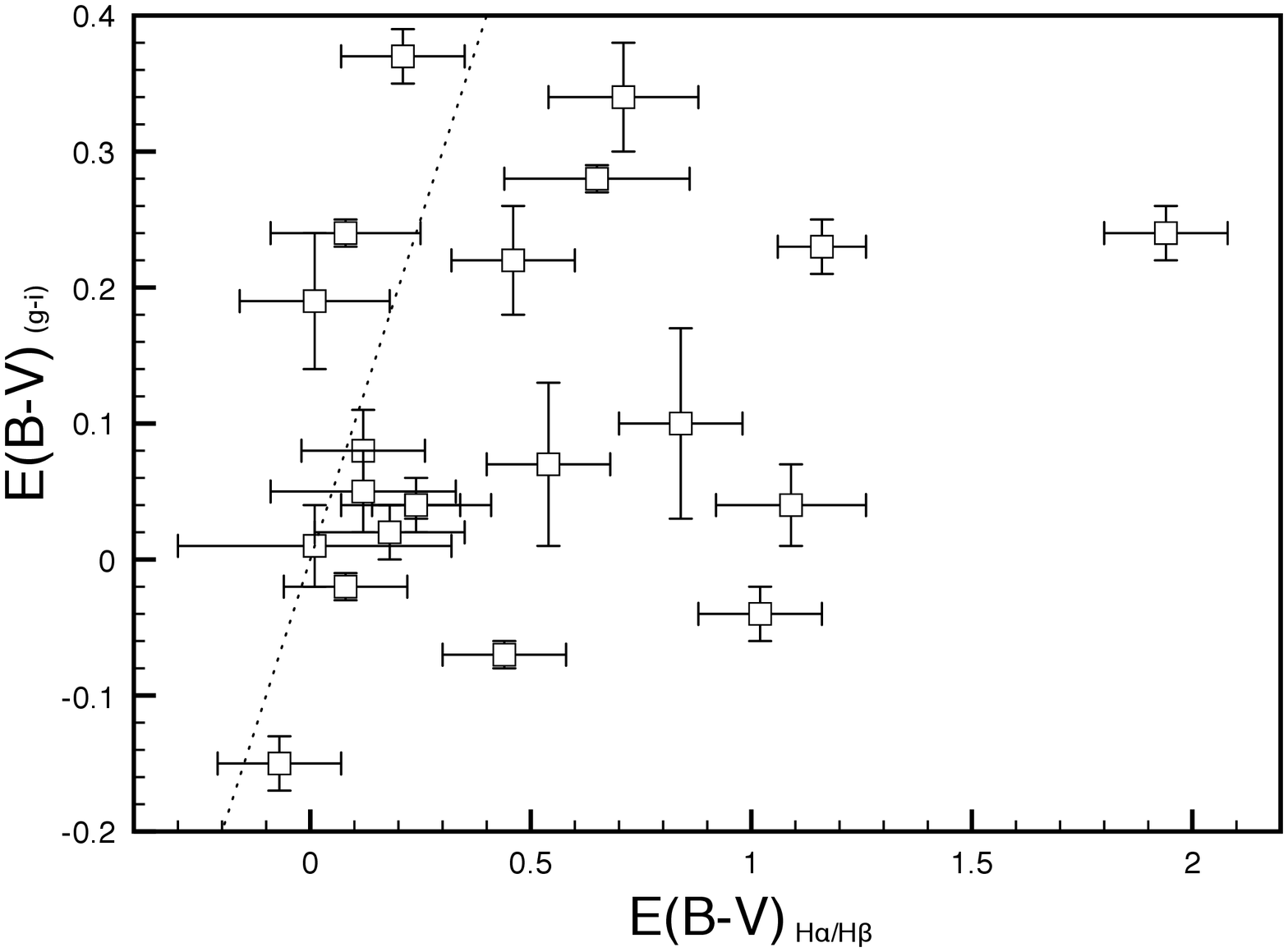}
}
\subfigure[]
{\label{fig-cana}
\includegraphics[trim = 0mm 0mm 0mm 0mm, clip, width=0.45\textwidth]{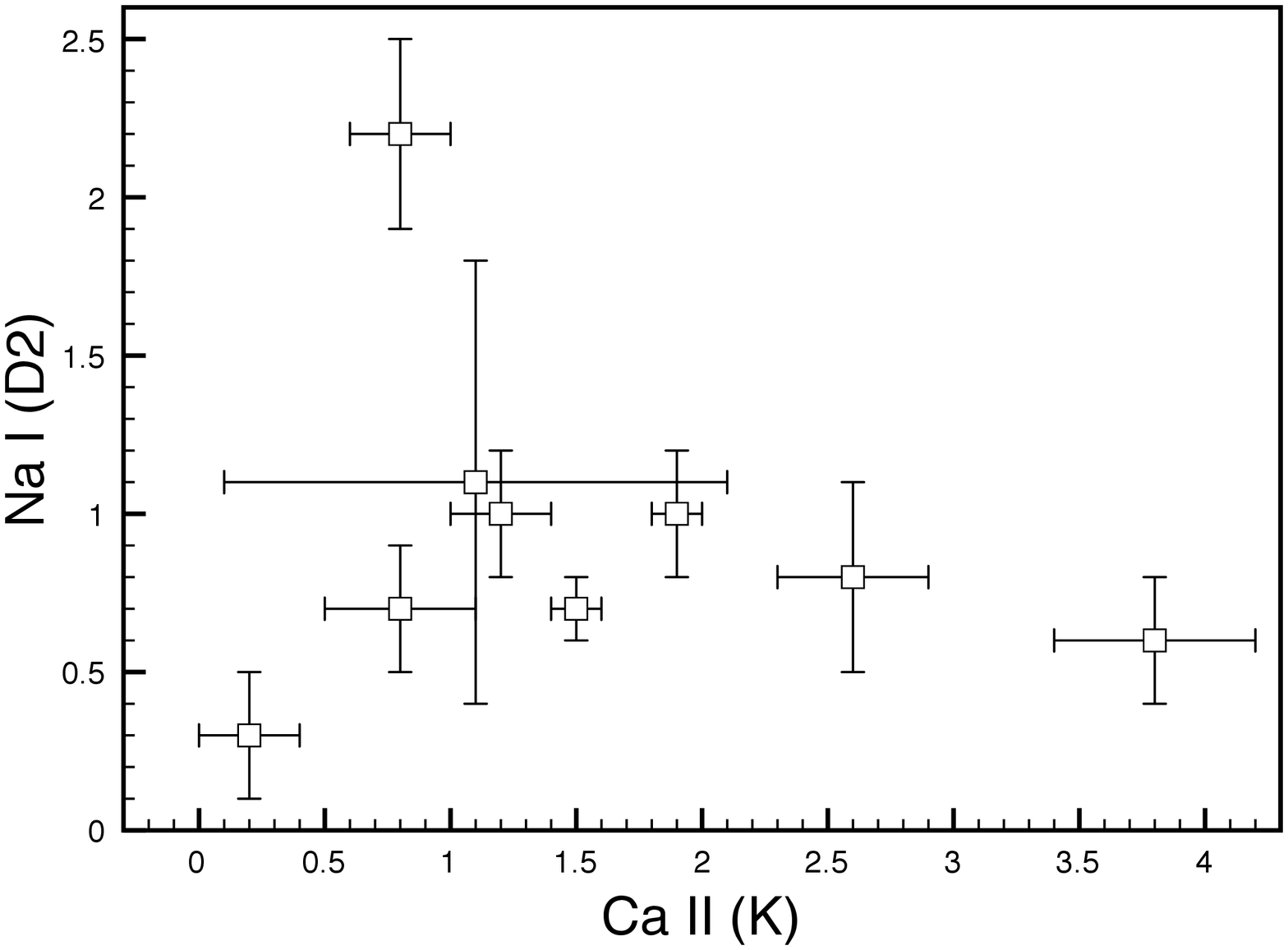}
}
\subfigure[]
{\label{fig-cana_b}
\includegraphics[trim = 0mm 0mm 0mm 0mm, clip, width=0.45\textwidth]{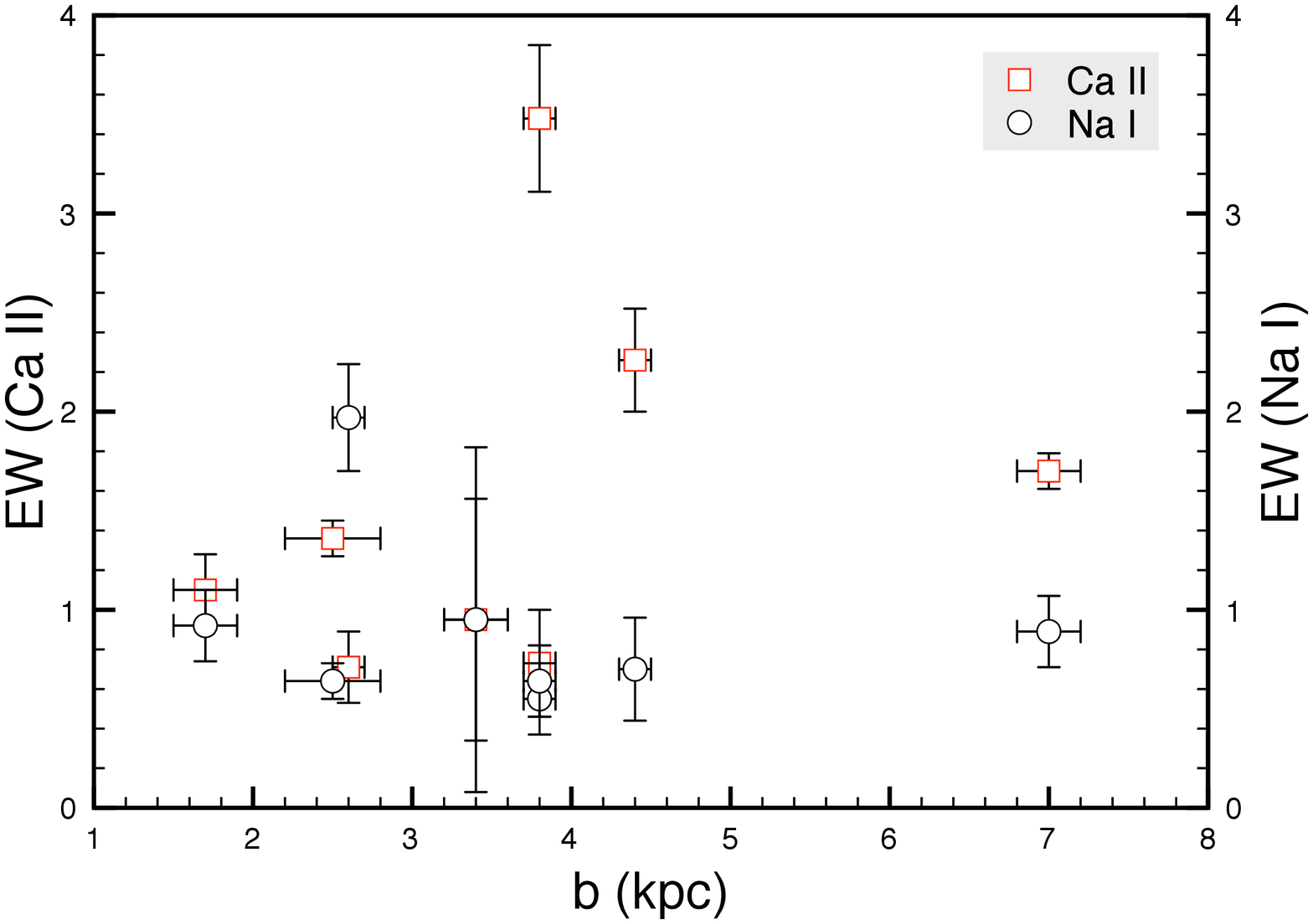}
}
\subfigure[]
{\label{fig-cana_ebv}
\includegraphics[trim = 0mm 0mm 0mm 0mm, clip, width=0.45\textwidth]{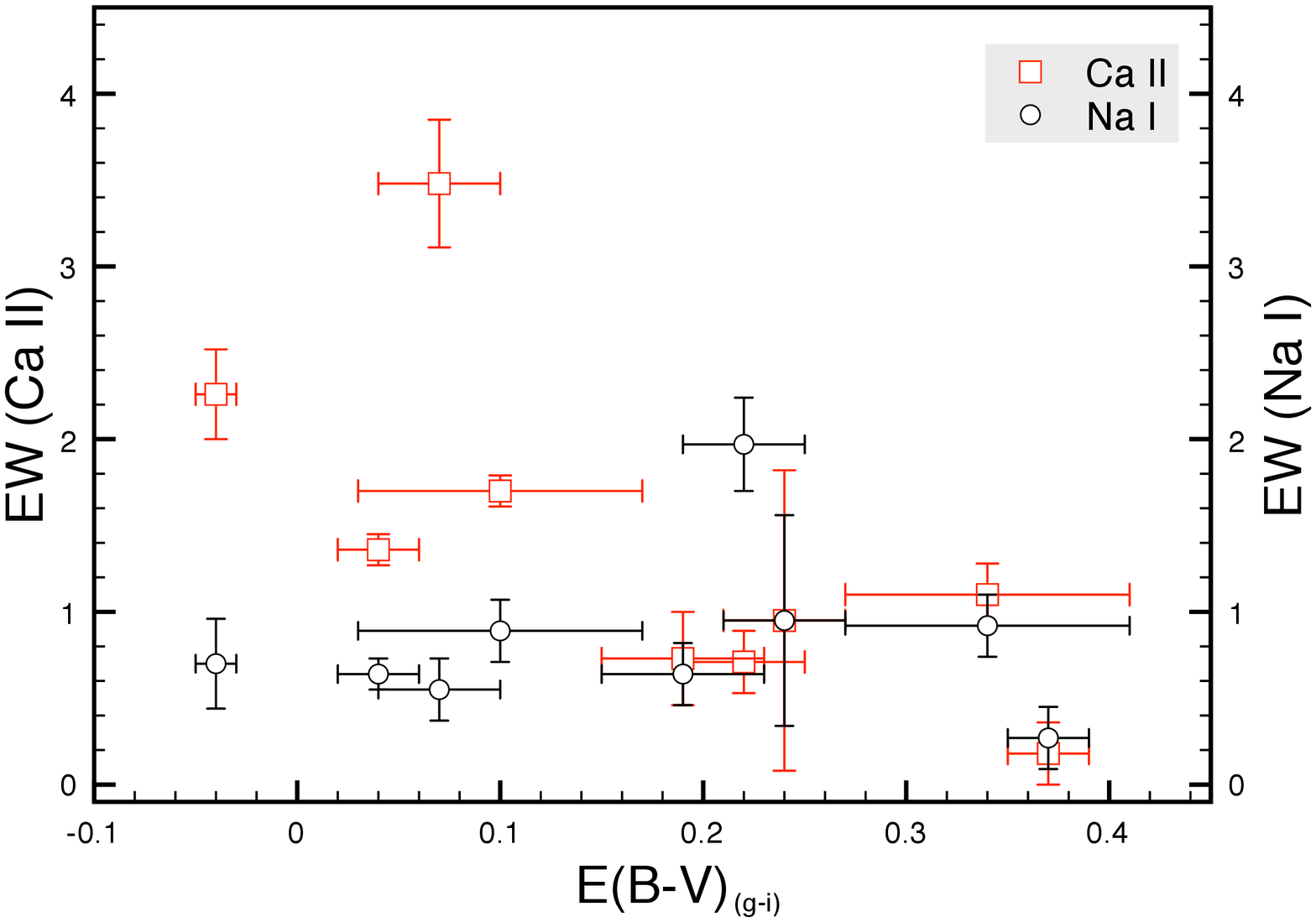}
}

\end{center}
\caption{a) Absorber rest-frame color excess vs. impact parameter and color. b) Histogram plotting the distribution of H$\alpha$/H$\beta$ in our sample. Overplotted are the points for absorber rest-frame color excess vs. H$\alpha$/H$\beta$. c) Absorber rest-frame color excess vs. reddening along the line of sight to the QSO. d) W(Na I) equivalent widths vs. W(Ca II) equivalent widths. e) W(Na I) and W(Ca II) equivalent widths vs. impact parameter. f) W(Na I) and W(Ca II) vs. absorber rest-frame color excess. }
\end{minipage}
\end{figure}


\begin{figure}
\begin{minipage}{1.0\linewidth}
\begin{center}

\includegraphics[trim = 0mm 0mm 0mm 0mm, clip, width=0.7\textwidth]{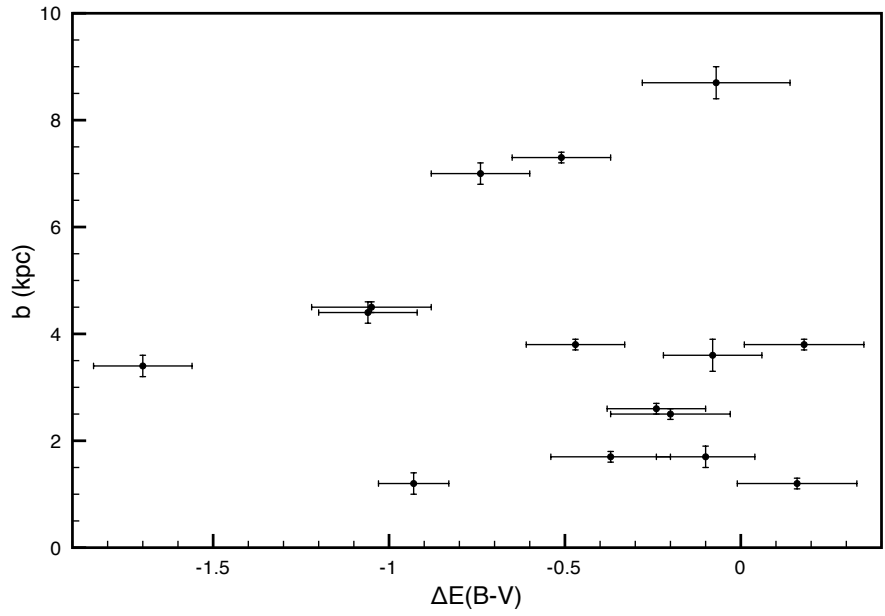}

\end{center}
\caption{Plot of $\Delta$E(B-V)$=$E(B-V)$_{(g-i)}$-E(B-V)$_{H\alpha/H\beta}$ vs. impact parameter. }\label{fig-ebvvsb}
\end{minipage}
\end{figure}


\begin{figure}
\begin{minipage}{1.0\linewidth}
\begin{center}

\includegraphics[trim = 0mm 0mm 0mm 0mm, clip, width=0.6\textwidth]{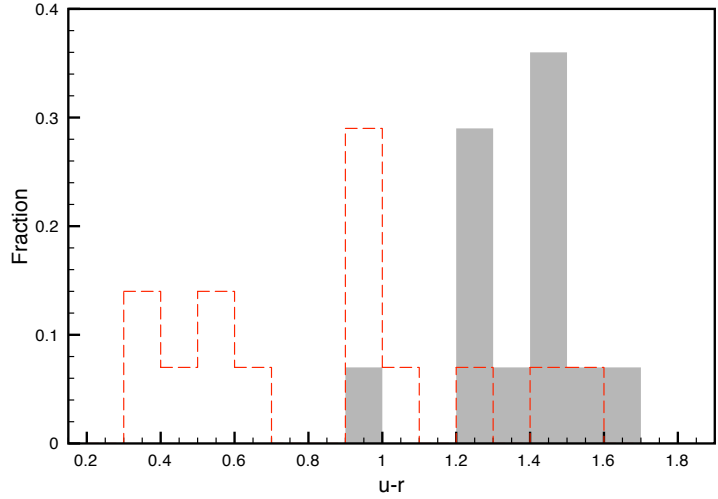}

\end{center}
\caption{Analogous plot for our data to Figure 8 of Strateva et al. (2001). The filled histogram corresponds to the fraction of the total sample in each bin, here 14 galaxies from our study which have (u-r) calculations. While our sample is too small to show the color separation between late- and early-type galaxies, we can compare it to the plot of Strateva et al. and see that our bins fall in the middle of the late-type range. The red dashed-line histogram corresponds to the (u-r) values from SDSS magnitudes for the same 14 galaxies prior to deblending the galaxy and quasar. See text for further explanation.}
\label{fig-color}
\end{minipage}
\end{figure}


\begin{figure}
\begin{minipage}{1.0\linewidth}
\begin{center}

\includegraphics[trim = 0mm 0mm 0mm 0mm, clip, width=0.7\textwidth]{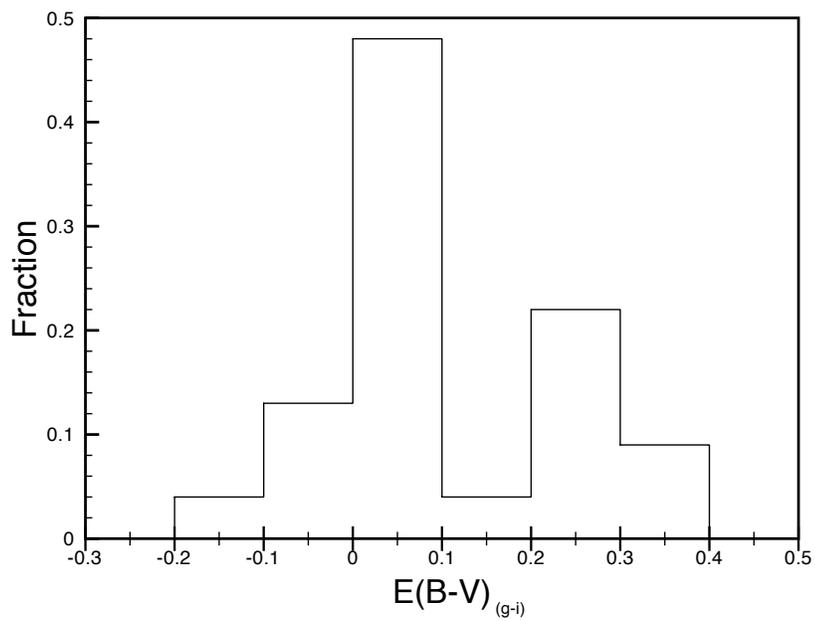}

\end{center}
\caption{
The histogram corresponds to the fraction of the total sample in each bin.  It's analogous to Figure 5 from York et al. (2006). }\label{fig-extinction}
\end{minipage}
\end{figure}


\begin{table}
\centering
\begin{minipage}{200mm}
\caption{Quasar galaxy pairs in SDSS DR5 found from galactic emission superimposed on the quasar spectrum.\label{tbl-qgp} }
\begin{tabular}{rlcccccc}
\\
\hline

Index$^{a}$ & Quasar Name & Plate & Fiber & MJD & $z_{QSO}$ & $z_{gal}$ & Quality \\

\hline

3 & SDSS J001342.44-002412.6 & 389 & 274 & 51795 & 1.6409 & 0.1557 & A \\

10 & SDSS J001637.08+135653.4 & 752 & 43 & 52251 & 2.5735 & 0.0912 & A \\

4 & SDSS J023914.39-070557.1$^{b}$ & 455 & 562 & 51909 & 0.7145 & 0.3423 & A \\

23 & SDSS J094041.69+341535.1$^{b}$ & 1594 & 48 & 52992 & 1.7178 & 0.4469 & B \\

1 & SDSS J094335.61-004322.0$^{bc}$ & 266 & 125 & 51630 & 0.2700 & 0.0989 & A \\

12 & SDSS J100514.20+530240.0 & 903 & 503 & 52400 & 0.5608 & 0.1358 & A \\

2 & SDSS J101015.74-010038.0$^{cd}$ & 270 & 205 & 51909 & 0.2300 & 0.0213 & C \\

14 & SDSS J101105.67+061917.2 & 996 & 119 & 52641 & 1.9496 & 0.0881 & A \\

15 & SDSS J104257.58+074850.5$^{b}$ & 1000 & 522 & 52643 & 2.6659 & 0.0335 & A \\

13 & SDSS J110224.13+573512.9$^{b}$ & 950 & 241 & 52378 & 1.6201 & 0.2932 & A \\

16 & SDSS J110735.68+060758.6 & 1004 & 281 & 52723 & 0.3801 & 0.1545 & A \\

11 & SDSS J114340.95+520303.3$^{b}$ & 881 & 313 & 52368 & 1.8164 & 0.1324 & A \\

19 & SDSS J123844.79+105622.2 & 1233 & 534 & 52734 & 1.3039 & 0.1185 & A \\

5 & SDSS J123846.69+644836.5 & 600 & 37 & 52317 & 1.5587 & 0.1190 & A \\

22 & SDSS J133733.23+445129.7$^{b}$ & 1465 & 300 & 53082 & 1.1678 & 0.1592 & A \\

20 & SDSS J140159.74+414156.2$^{be}$ & 1346 & 60 & 52822 & 1.7005 & 0.1202 & A \\

17 & SDSS J143458.04+504118.6$^{b}$ & 1046 & 542 & 52460 & 1.4850 & 0.1992 & A \\

18 & SDSS J145240.53+544345.1$^{b}$ & 1163 & 505 & 52669 & 1.5195 & 0.1026 & A \\

21 & SDSS J145938.49+371314.7 & 1352 & 355 & 52819 & 1.2189 & 0.1489 & A \\

6 & SDSS J150140.45+571026.1$^{b}$ & 610 & 133 & 52056 & 1.7993 & 0.1037 & A \\

8 & SDSS J160248.50+491640.7 & 622 & 340 & 52054 & 1.0734 & 0.2061 & D \\

7 & SDSS J160521.27+510740.8$^{bf}$ & 620 & 535 & 52375 & 1.2290 & 0.0994 & A \\

9 & SDSS J161016.17+500728.8$^{bc}$ & 623 & 304 & 52051 & 0.2400 & 0.1267 & A \\

\hline

\end{tabular}

\footnotetext[1]{Index number are arranged by increasing plate number in SDSS.}
\footnotetext[2]{Systems also found by Noterdaeme et al. (2010).}
\footnotetext[3]{Seyfert classification with line width $<1000$ km/s.}
\footnotetext[4]{Galaxy name: SDSS J101015.55-010038.7}
\footnotetext[5]{Galaxy name: SDSS J140159+414154.5}
\footnotetext[6]{Galaxy name: SDSS J160521.29+510738.7}

\end{minipage}
\end{table}


\begin{table}
\centering
\begin{minipage}{200mm}
\caption{QSO offsets and galaxy sizes (measured in SDSS r-band)}\label{tbl-offset}
\begin{tabular}{ccccccccccc}
\\
\hline

Index & $\Delta$RA & $\Delta$Dec & $\theta$ & b & Length  & Width & Maj. Axis \\
  & (pixels) & (pixels) & (\arcsec) & (kpc) & (\arcsec) & (\arcsec) & (degrees) & \\

\hline

3  & 6.3 & 7.0 & 1.3 & 3.4 & 9.5 & 2.7 & 40  \\
10 & 9.4 & 10.8 & 2.3 & 3.8 & 10.8 & 9.4 & 0  \\
4  & --$^{a}$ & -- & -- & -- & -- & -- & --   \\
23  & -- & -- & -- & -- & --  & -- & --  \\
1  & 5.0 & 5.9 & 1.4 & 2.5 & 6.0 & 4.5 & 45  \\
12  & 11.8 & 5.4 & 1.5 & 3.6 & 13.0 & 4.1 & 50  \\
2  & 9.1 & 11.5 & 2.9 & 1.2 & 11.5 & 9.1 & 0  \\
14 & 17.1 & 4.4 & 1.1 & 1.7 & 17.7 & 3.5 & 80  \\
15  & 14.3 & 13.6 & 2.7 & 1.7 & 14.3 & 13.6 & 0  \\
13  & -- & -- & -- & -- & -- & -- & --  \\
16  & 3.5 & 2.4 & 1.7 & 4.5 & 3.5 & 2.4 & 90  \\
11  & -- & -- & -- & -- & -- & -- & --  \\
19  & 7.8 & 9.6 & 1.2 & 2.6 & 9.6 & 7.8 & 0  \\
5  & 9.0 & 4.0 & 3.3 & 7.0 & 9.8 & 5.6 & -45  \\
22  & -- & -- & -- & -- & -- & -- & --  \\
20  & 3.6 & 3.2 & 0.5 & 1.2 & 3.6 & 3.2 & 0  \\
17  & 7.6 & 10.3 & 2.3 & 7.3 & 10.3 & 7.6 & 0  \\
18 & -- & -- & -- & -- & -- & -- & --  \\
21  & 6.9 & 6.5 & 1.7 & 4.4 & 9.5 & 3.5 &45  \\
6  & 14.8 & 6.7 & 4.6 & 8.7 & 14.8 & 6.8 & 0  \\
8  & -- & -- & -- & -- & -- & -- & --  \\
7  & 10.3 & 6.2 & 2.1 & 3.8 & 10.4 & 6.1 & 90  \\
9  & -- & -- & -- & -- & -- & -- & --  \\

\hline

\end{tabular}
\footnotetext[1]{-- means that no offset could be determined.}
\end{minipage}
\end{table}

\begin{table}
\centering
\begin{minipage}{200mm}
\caption{Photometric data for deconvolved quasars and galaxies\label{tbl-phot}}
\begin{tabular}{ccccccccccccccc}
\\
\hline

Index & i mag & (g-i) & $\Delta$(g-i)& E(B-V)$_{(g-i)}$ & m$_{r}$ & (u-r) & M$_{r}$  & L$^{*}_{r}$ & M$_{dev(r)}$ & M$_{exp(r)}$ & M$_{psf(r)}$ & Petrosian  \\
&Quasar &Quasar & & & & galaxy & galaxy & galaxy & & &  & \arcsec  \\

\hline

3 & 18.47 & 0.72 & 0.43 & 0.24  & 18.03 & 1.49 & -21.38 & 1.69 & 17.51$^{a}$ & 17.77 & 18.58 &3.16  \\
10 & 19.68 & 0.37 & 0.12 & 0.07  & 17.41 & 1.43 & -20.69 & 0.90 & 16.87 & 17.31$^{a}$ & 19.11 &4.83  \\
4 & 19.28 & 0.05 & 0.01 & 0.01  & -- & -- & --  & -- & 19.32$^{a}$ & 19.34 & 19.41 &1.92 \\
23 & 19.05 & 0.31 & 0.02 &0.01  & -- & -- & -- & --  & 19.23$^{a}$ & 19.23 & 19.24 &1.17 \\
1 & 19.08 & 0.65 & 0.07 & 0.04  & 19.64 & 1.64 & -18.69 & 0.14 & 18.67$^{a}$ & 18.80 & 19.06 &2.33 \\
12 & 19.01 & -0.16 & -0.26 & -0.15  & 19.50 & 1.28 & -19.55 & 0.31 & 18.56$^{a}$ & 18.65 & 18.79 &1.97 \\
2 & 18.87 & 1.11 & 0.35 & 0.23  & 17.91 & 1.47 & -16.87 & 0.03 & 18.87$^{a}$ & 19.10 & 19.39 &3.10\\
14 & 18.88 & 0.81 & 0.57 & 0.34  & 18.23 & 1.25 & -19.78 & 0.39 & 17.85$^{a}$ & 18.23 & 19.18 &4.27 \\
15 & 19.03 & 0.23 & -0.02 & -0.02  & 17.93 & 1.58 & -17.87 & 0.07 & 17.49$^{a}$ & 17.86 & 18.88 &6.22 \\
13 & 19.11 & 0.36 & 0.07 & 0.04  & -- & -- & -- & -- & 19.30$^{a}$ & 19.30 & 19.35 &1.40\\
16 & 19.19 & 0.29 & 0.07 & 0.04  & 20.16 & 1.32 & -19.20 & 0.23 & 18.89$^{a}$ & 18.96 & 19.17 &1.93\\
11 & 17.03 & 0.96 & 0.65 & 0.37  & -- & -- & -- & --  & 17.39 & 17.35$^{a}$ & 17.39 &1.64 \\
19 & 18.96 & 0.67 & 0.39 & 0.22  & 18.46 & 1.47 & -20.29 & 0.62 & 17.97$^{a}$ & 18.20 & 18.75 &3.76\\
5 & 17.70 & 0.47 & 0.17 & 0.10  & -- & -- & -- & --  & 17.63$^{a}$ & 17.68 & 17.93 &2.27 \\
22 & 18.66 & 0.40 & 0.14 & 0.08  & -- & -- & -- & -- & 18.64$^{a}$ & 18.64 & 18.66 &1.16 \\
20 & 17.94 & 0.70 & 0.41 &0.24  & 19.95 & 1.25 & -18.80 & 0.16 & 19.53 & 19.95$^{a}$ & 21.39 &7.36 \\
17 & 19.22 & 0.21 & -0.12 & -0.07  & 18.91 & 0.95 & -21.04  &1.24 & 18.33$^{a}$ & 18.54 & 19.14 &3.85 \\
18 & 18.41 & 0.36 & 0.03 & 0.02  & -- & -- & -- & -- & 18.43$^{a}$ & 18.48 & 18.63 &1.94\\
21 & 19.14 & 0.23 & -0.07 & -0.04  & 19.67 & 1.24 & -18.37 & 0.11 & 18.68$^{a}$ & 18.85 & 19.02 &2.13 \\
6 & 18.72 & 0.33 & 0.08 & 0.05  & 19.26 & 1.43 & -19.15  & 0.22 & 18.51$^{a}$ & 18.72 & 18.81 &3.42 \\
8 & 19.24 & 0.23 & 0.05 & 0.03  & -- & -- & --  &-- & 19.17$^{a}$ & 19.17 & 19.18 &1.43 \\
7 & 18.41 & 0.58 & 0.33 & 0.19  & 19.06 & 1.94 & -19.30 & 0.25 & 18.53$^{a}$ & 18.54 & 18.58 &1.25 \\
9 & 18.13 & 1.07 & 0.49 & 0.28  & -- & -- & --  &-- & 18.48$^{a}$ & 18.64 & 18.95 &1.80 \\

\hline

\end{tabular}
\footnotetext[1]{Indicates that the best fit profile in the r-band image.}
\end{minipage}
\end{table}


\begin{table}\tiny
\centering
\begin{minipage}{200mm}
\caption{Emission line strengths for galaxies in front of quasars.}
\label{tbl-emission}
\begin{tabular}{lccccccccccccccccccccccccc}
\\
\hline

ID & H$\alpha$ & $f_{H\alpha}$$^{a}$  & H$\beta$ & $f_{H\beta}$  & [O II]  & $f_{[O II]}$  & [O III]a & $f_{[O III]a}$ & [O III]b & $f_{[O III]b}$  & N IIa & $f_{N IIa}$ & N IIb & $f_{N IIb}$ & [S II]a  & $f_{[S II]a}$ & [S II]b & $f_{[S II]b}$ \\
& $\lambda_{obs}$$^{b}$& & $\lambda_{obs}$&  & $\lambda_{obs}$ &  & $\lambda_{obs}$ &  &$\lambda_{obs}$ & &$\lambda_{obs}$ & &$\lambda_{obs}$ & & $\lambda_{obs}$ & & $\lambda_{obs}$ & &\\

\hline

3	&7584.4	&139.0	&5619.7	&8.2	&4308.4	&--	&5734.1	&--$^{c}$ &5788.2 & 14.0 & 7569.3 & 13.9 &7612.1	&24.1	&7762.5	&35.2	&7780.2	&28.2	\\
10	&7161.2	&241.0	&5304.4	&51.6	&4066.9	&106.1	&5411.8	&13.4 &5463.5 & 25.8 & 7145.2 & 31.1 &7183.7	&102.0	&7329.1	&46.2	&7343.9	&39.0		\\
4	&8808.1	&21.6	&6525.0	&7.4 	&5002.9	&20.9	&6656.0	&11.1 & 6719.9 & 37.3 & 8789.9 & -- &8837.5	&--	&9015.9	&--	&9035.3	&--	\\
23	&9493.5	&--	&7034.0	&9.0  &--  &--	&7175.4	&11.8 & 7244.7 & 20.6 & 9471.2 & -- &9522.3	&--	&9714.7	&--	&9735.5	&--	\\
1	&7210.7	&61.5	&5241.5	&17.0	&4095.8	&102.0	&5448.6	&14.5 & 5501.2 & 34.1 & 7195.1 & -- &7234.2	&13.6	&7380.1	&12.9	&7395.2	&8.9	\\
12	&7454.3	&55.8	&5522.0	&20.6	&4233.9	&98.6	&5634.9	&--  & 5686.9 & 34.4 & 7440.7 & -- &7478.6	&7.3	&7631.9	&--	&7648.4	&--\\
2	&6702.6	&39.0	&4964.7	&4.7	&3805.3	&--	&5065.5	&11.2 & 5113.5 & 33.1 & 6687.5 & --	 &6723.6	&--	&6859.4	&--	&6874.1	&--	\\
14	&7140.9	&76.0	&5290.2	&13.8	&4055.4	&23.6	&5396.8	&-- & 5447.7 & 15.7	& 7124.1 & 9.0 &7162.4	&28.9	&7307.8	&19.3	&7323.5	&8.8 \\
15	&6782.7	&74.6	&5023.8	&24.2	&3853.1	&53.0	&5124.9	&18.6 & 5174.5 & 42.5 & 6766.1 & -- &6803.0	&8.0	&6941.6	&16.0	&6955.8	&15.3	\\
13	&8487.3	&66.5	&6287.5	&18.6	&4820.1	&23.9	&6412.8	&9.7 	 & 6475.2 & 44.1 & 8469.0 & -- &8514.8	&--	&8686.8	&--	&8705.4	&--\\
16	&7576.8	&78.7	&5613.6	&10.0	&4303.3	&29.1	&5729.2	&-- & 5781.2 & 24.5	& 7561.2 & 9.4 &7605.7	&24.4	&7757.5	&3.4	&7776.3	&-- \\
11	&7431.8	&121.0	&5505.5	&34.8	&4221.5	&78.1	&5615.6	&74.2 & 5670.1 & 179.0 & 7414.5 & -- &7454.6	&--	&7605.1	&--	&7621.4	&--	\\
19	&7340.8	&121.0	&5437.6	&27.7	&4169.5	&87.1	&5550.6	&-- & 5600.4 & 22.8 	& 7324.6 & 17.3 &7363.7	&40.6	&7512.9	&27.7	&7529.2	&27.5 \\
5	&7343.5	&115.0	&5439.5	&18.4	&4170.7	&39.8	&5550.6	&-- & 5604.2 & -- & 7328.6 & 31.5 &7366.2	&60.4	&7515.7	&20.6	&7529.2	&21.0	\\
22	&7607.7	&49.2	&5635.6	&15.5	&4321.6	&58.7	&5748.2	&26.9 & 5803.9 & 67.6  & 7591.3 & -- &7632.4	&--	&7786.5	&--	&7803.2	&--	\\
20	&7351.4	&223.0	&5445.5	&72.4	&4175.6	&220.0	&5554.6	&72.5 & 5608.4 & 196.0 & 7335.9 & -- &7374.4	&18.1	&7524.5	&43.6	&7539.6	&27.0		\\
17	&7869.9	&147.0	&5829.7	&34.5	&4469.9	&84.0	&5947.4	&13.9 & 6004.4 & 48.3 & 7854.2 & --	 &7896.2	&22.8	&8054.0	&27.6	&8071.4	&37.0\\
18	&7236.4	&78.7	&5360.7	&23.4	&4109.9	&86.3	&5467.9	&18.8 &5520.8 & 50.7 &7218.0 & -- &7257.0	&--	&7405.6	&16.1	&7421.9	&10.7	\\
21	&7539.8	&75.6	&5585.4	&10.3	&4282.6	&42.4	&5699.4	&-- &5752.9 & 44.4 & 7525.8 & -- &7563.7	&16.7	&7716.8	&25.3	&7733.4	&24.8	\\
6	&7243.3	&78.1	&5365.4	&24.6	&4113.9	&96.3	&5472.9	&22.0 & 5525.8 & 53.8 &7231.1 & -- &7265.5	&11.5	&7412.5	&32.5	&7428.2	&18.7\\
8	&7915.6	&35.4	&5863.1	&--	&4494.8	&--	&5982.1	&-- & 6039.9 & -- & 7899.2 & -- &7941.9	&--	&8102.3	&--	&8119.6 &--	\\
7	&7214.9	&96.1	&5344.4	&32.8	&4097.8	&104.0	&5452.3	&27.2 & 5504.2 & 57.5 & 7200.6 & 10.0  &7238.1	&25.7	&7383.8	&24.6	&7400.0	&19.9 	\\
9	&7394.1	&34.3	&5478.2	&6.6		&4200.6	&19.3	&5587.3	&12.0  & 5641.4 & 60.1 & 7381.7 & -- &7421.6	&--	&7571.5	&--	&7587.8	&--\\

\hline

\end{tabular}
\footnotetext[1]{Flux measured in units of 10$^{-17}$ ergs cm$^{-2}$  s$^{-1}$. }
\footnotetext[2]{Observed wavelengths measured in angstroms.}
\footnotetext[3]{Indicates no emission line detected or redshifted out of range.}
\end{minipage}
\end{table}


\begin{table}
\centering
\begin{minipage}{200mm}
\caption{Spectroscopic properties of galaxies\label{tbl-SFR}}
\begin{tabular}{cccccccccc}
\\
\hline

Index & H$\alpha$/H$\beta$ & E(B-V)$_{H\alpha/H\beta}$ &SFR$_{H\alpha}$$^{a}$ & SFR$_{[O II]}$$^{a}$ & SFR$_{H\alpha}$$^{b}$ & SFR$_{[O II]}$$^{b}$ \\
& & & M$_{\sun}$ yr$^{-1}$ & M$_{\sun}$ yr$^{-1}$ & M$_{\sun}$ yr$^{-1}$ & M$_{\sun}$ yr$^{-1}$\\

\hline

3 & 16.9 & 1.94 & 0.77 & -- & 60.3 &-- \\
10 & 4.7 & 0.54  & 0.40 & 0.15 & 1.33 & 1.34  \\
4 & 2.9 & 0.01 & 0.66 & 0.53 & 0.67 & 0.55  \\
23 & -- & -- & -- & -- &-- &--  \\
1 & 3.6 & 0.24 & 0.12 & 0.17 & 0.21 &0.47 \\
12 & 2.7 & -0.07 & 0.22 & 0.32 & 0.19 & 0.24 \\
2 & 8.3 & 1.16 & 0.003  & 0.003 & 0.04 & 0.41  \\
14 & 5.5 & 0.71 & 0.12 & 0.03 & 0.57 & 0.57 \\
15 & 3.1 & 0.08 & 0.02  & 0.01 & 0.02 & 0.01 \\
13 & 3.6 & 0.24 & 1.44 & 0.43 & 2.49 & 1.19 \\
16 & 7.8 & 1.09 & 0.38 & 0.12 & 4.41 & 10.8  \\
11 & 3.5 & 0.21 & 0.44 & 0.24 & 0.72 & 0.58 \\
19 & 4.4 & 0.46 & 0.35 & 0.21 & 1.00 & 1.45 \\
5 & 6.2 & 0.84 & 0.34 & 0.10 & 2.21 & 3.16 \\
22 & 3.2 & 0.12 & 0.27 & 0.27  & 0.35 & 0.43 \\
20 & 3.1 & 0.08 & 0.66 & 0.54 & 0.80 & 0.76  \\
17 & 4.3 & 0.44 & 1.32 & 0.63 & 3.54 & 3.89 \\
18 & 3.4 & 0.18 & 0.17 & 0.15 & 0.25 & 0.32  \\
21 & 7.3 & 1.02 & 0.36 & 0.17 & 3.55 & 11.5 \\
6 & 3.2 & 0.12 & 0.17 & 0.17 & 0.22 & 0.28 \\
8 & -- & -- & 0.36 & -- & -- &--  \\
7 & 2.9 & 0.01 & 0.19 & 0.17 & 0.19 & 0.18 \\
9 & 5.2 & 0.65  & 0.12 & 0.06 & 0.52 & 0.83 \\

\hline

\end{tabular}
\footnotetext[1]{Uncorrected for extinction.}
\footnotetext[2]{Corrected for extinction.}
\end{minipage}
\end{table}


\begin{table}
\centering
\begin{minipage}{200mm}
\caption{SDSS Deblended Galaxy Magnitudes}\label{tbl-deblended}
\begin{tabular}{ccccccccccccc}
\\
\hline

Index & m$_{u}$ & S$^{a}$ m$_{u}$  & m$_{r}$ & S m$_{r}$   & (u-r) & S (u-r) \\

\hline

2	& 19.38	& 19.39		& 17.91	& 17.91	& 1.47	& 1.48	\\	
7	& 21.00	& 20.98		& 19.06	& 19.05	& 1.94	& 1.93	\\
20	& 21.20	& 21.23		& 19.95	& 19.96	& 1.25	& 1.27	\\

\hline

\end{tabular}
\footnotetext[1]{S indicates the SDSS values for the filter.}
\end{minipage}
\end{table}


\begin{table}
\centering
\begin{minipage}{200mm}
\caption{Rest-frame equivalent widths (W) for Ca II and Na I interstellar absorption lines in foreground galaxies.}\label{tbl-ew}
\begin{tabular}{ccccccccc}
\\
\hline

Index & Ca II K & Ca II K  & Ca II H & Ca II H  & Na I D2 & Na I D2 & Na I D1 & Na I D1 \\
 & $\lambda$3934 & W$_{obs}$$^{a}$ (\AA) & $\lambda$3969 & W$_{obs}$ (\AA) & $\lambda$5890 & W$_{obs}$$^{a}$ (\AA) & $\lambda$5896 & W$_{obs}$ (\AA) \\

\hline

3	& 4546.8 	& 1.0 	& 4588.1 	& 1.6 	& 6807.4	& 1.0 	& 6814.4 	& 1.0\\
10 	& 4291.1 	& 3.5  	& 4326.9 	& 6.7 	& 6428.3 	& 0.6 	& 6434.6 	& 0.6\\
4 	& 5277.1 	& --  		& 5325.4	& -- 		& 7903.0 	& --   		& 7911.0 	& --\\
23 	& 5690.9 	& 0.6 	& 5741.2 	& -- 		& 8520.0 	& --    	& 8528.7 	& --\\
1 	& 4324.0 	& 1.4   	& 4361.8 	& 1.8 	& 6471.5 	& 0.6   	& 6476.5 	& --\\
12 	& 4468.5 	& 0.5 	& 4506.8 	& -- 		& 6688.1 	& --   		& 6694.9 	& --\\
2 	& 4016.2 	& --  		& 4052.9 	& -- 		& 6014.5 	& --   		& 6020.7 	& --\\
14 	& 4280.6 	& 1.1   	& 4318.4 	& 0.5 	& 6408.7 	& 0.9  	& 6415.2 	& 0.2\\
15 	& 4063.6 	& --  		& 4100.8 	& -- 		& 6085.6 	& --    	& 6091.8 	& --\\
13 	& 5084.9 	& --  		& 5131.5 	& -- 		& 7615.1 	& --   		& 7622.9 	& --\\
16 	& 4539.5 	& --   		& 4581.0 	& -- 		& 6798.3 	& --    	& 6805.2  & --\\
11 	& 4453.7 	& 0.2  	& 4500.4 	& 0.5 	& 6667.5 	& 0.3    	& 6676.1 	& 0.3\\
19 	& 4399.6 	& 0.7   	& 4440.8 	& 3.2 	& 6588.8 	& 2.0   	& 6597.6 	& 1.3\\
5 	& 4401.1 	& 1.7  	& 4440.5 	& 1.3 	& 6591.3 	& 0.9   	& 6598.3 	& 0.4\\
22 	& 4557.9 	& --   		& 4599.6 	& -- 		& 6825.9 	& --    	& 6832.3 	& 0.6\\
20 	& 4404.3 	& --    	& 4444.7 	& -- 		& 6595.9 	& --    	& 6602.6 	& --\\
17 	& 4715.8 	& 0.9   	& 4758.2 	& -- 		& 7061.2 	& --   		& 7068.4 	& --\\
18 	& 4335.4 	& --   		& 4375.1 	& -- 		& 6496.0 	& 0.4    	& 6501.4 	& 1.6 \\
21 	& 4518.9 	& 2.3     	& 4560.4 	& 2.2 	& 6766.8 	& 0.7   	& 6774.1 	& 0.8\\
6 	& 4339.6 	& --  		& 4379.3 	& -- 		& 6499.0 	& --   		& 6505.6 	& --\\
8 	& 4742.2 	& --   		& 4785.6 	& -- 		& 7101.8 	& --    	& 7109.1 	& --\\
7 	& 4323.3 	& 0.7    	& 4362.2 	& -- 		& 6475.6 	& 0.6     	& 6480.9 	& 0.4\\
9 	& 4430.0 	& --  		& 4470.5 	& -- 		& 6634.3 	& --   		& 6641.0 	& --\\

\hline

\end{tabular}
\footnotetext[1]{Rest-frame equivalent widths were measured in the original SDSS spectra using IRAF. }
\end{minipage}
\end{table}

\bsp

\label{lastpage}

\end{document}